\documentclass[aps,prx,twocolumn,superscriptaddress,showpacs,longbibliography]{revtex4-2}

\usepackage{amsmath,graphicx}
\usepackage[T1]{fontenc}
\usepackage{xcolor}
\usepackage{gensymb}
\usepackage{siunitx}
\usepackage{ulem}
\usepackage{color}
\usepackage{multirow}
\usepackage{bm}
\usepackage{annotate-equations}
\IfFileExists{newtxtext.sty}
{\usepackage{newtxtext,newtxmath}}
{\IfFileExists{stix.sty}
	{\usepackage{stix}}
	{\IfFileExists{mathptmx.sty}
		{\usepackage{mathptmx}}{} } }

\usepackage{color}
\definecolor{LinkColor}{rgb}{0.256,0.439,0.588}
\usepackage{hyperref}
\usepackage{orcidlink}
\usepackage{comment}
\hypersetup{
	pdfauthor={good guys},
	pdftitle={good title},
	colorlinks=true,
	citecolor=LinkColor,
	linkcolor=LinkColor,
	urlcolor=LinkColor
}

\def\bk{{\mathbf{k}}}
\def\bK{{\mathbf{K}}}
\def\bR{{\mathbf{R}}}

\def\bq{{\mathbf{q}}}
\def\bQ{{\mathbf{Q}}}
\def\bG{{\mathbf{G}}}
\def\bA{{\mathbf{A}}}
\def\mQ{{\mathcal{Q}}}
\def\G{{\mathbf{G}}}

\def\I{{\mathrm{I}}}
\def\e{{\mathrm{e}}}

\def\s{{\mathrm{s}}}
\def\m{{\mathrm{m}}}
\def\g{{\mathrm{g}}}
\def\r{{\mathrm{r}}}
\def\bbT{\mathbb{T}}
\def\T{\mathrm{T}}
\def\det{{\mathrm{det}}}
\def\Tr{{\mathrm{Tr}}}

\def\vc{\overrightarrow{c}}

\makeatletter
\renewcommand{\fnum@figure}{Fig. \thefigure}
\makeatother

\begin{document}
\title{Strain-Tuned Incommensurate Kekul\'e Spiral Order in Twisted Bilayer Graphene: \\A Quantum Many-Body Study}   

\author{Cheng Huang}
\thanks{These authors contributed equally to this work.}
\affiliation{Department of Physics and HK Institute of Quantum Science \& Technology, The University of Hong Kong, Pokfulam Road,  Hong Kong SAR, China}
\affiliation{State Key Laboratory of Optical Quantum Materials, The University of Hong Kong, Pokfulam Road,  Hong Kong SAR, China}

\author{Yves H. Kwan}
\thanks{These authors contributed equally to this work.}
\affiliation{Department of Physics, University of Texas at Dallas, Richardson, Texas 75080, USA}
\affiliation{Princeton Center for Theoretical Science, Princeton University, Princeton, NJ 08544}

\author{Maksim Ulybyshev}
\affiliation{Institut f\"ur Theoretische Physik und Astrophysik, Universit\"at W\"urzburg, 97074 W\"urzburg, Germany}

\author{Fakher F. Assaad \orcidlink{0000-0002-3302-9243}}
\email{fakher.assaad@uni-wuerzburg.de}
\affiliation{Institut f\"ur Theoretische Physik und Astrophysik, Universit\"at W\"urzburg, 97074 W\"urzburg, Germany}
\affiliation{W\"urzburg-Dresden Cluster of Excellence ctd.qmat, Am Hubland, 97074 W\"urzburg, Germany}

\author{Laura Classen}
\email{l.classen@fkf.mpg.de}
\affiliation{Max Planck Institute for Solid State Reserach, Heisenbergstr. 1, 70569 Stuttgart Germany}
\affiliation{School of Natural Sciences, Technische Universit\"at München, 85748 Garching, Germany}
\affiliation{Munich Center for Quantum Science and Technology (MCQST), Schelingstr. 4, D-80799 München, Germany}

\author{Zi Yang Meng \orcidlink{0000-0001-9771-7494}}
\email{zymeng@hku.hk}
\affiliation{Department of Physics and HK Institute of Quantum Science \& Technology, The University of Hong Kong, Pokfulam Road,  Hong Kong SAR, China}
\affiliation{State Key Laboratory of Optical Quantum Materials, The University of Hong Kong, Pokfulam Road,  Hong Kong SAR, China}

\begin{abstract}

The physics of twisted bilayer graphene away from the exactly solvable chiral limit and the quantum Monte Carlo sign-problem-free charge neutrality point is elusive due to the exponential increase in the computational complexity, which has rendered explanations of experimentally observed insulating and superconducting phases restricted largely to the perturbative level. Here we focus on the filling factor $\nu=\pm2$ and address the question of the strain dependence of the interacting ground state by approximate quantum Monte Carlo (AQMC), state-of-the-art exact diagonalization (ED) and Hartree-Fock (HF) mean field, in order to investigate the strain-tuned transition from the Kramers intervalley coherent (KIVC) state to the incommensurate Kekul\'e spiral (IKS) state. While all three methods capture the KIVC order, only ED and HF detect the weaker IKS order that AQMC does not capture adequately. As the AQMC is still capable of capturing stronger orders like the KIVC, our combined protocol may open the door for further understanding of the rich phases of twisted bilayer graphene and other strongly-correlated systems.
\end{abstract}
\date{\today}
\maketitle

\section{Introduction}
Understanding the phase diagram of twisted bilayer graphene (TBG) has remained an enduring pursuit in strongly-correlated moir\'e materials since the observation of superconductivity and correlated insulating behavior at the magic angle $\theta\simeq1.05^\circ$~\cite{Cao2018insulator,Cao2018superconductivity}. Beyond the twist angle, it has been recognized that strain considerably alters the physics of realistic TBG devices. In particular, uniaxial heterostrain, which is often present with strength $\epsilon_\s\sim0.1\%-0.7\%$~\cite{kerelsky2019maximized,choi2019electronic,xie2019spectroscopic,wong2020cascade,kazmierczak2021strain}, breaks $C_3$ rotation symmetry and significantly enhances the non-interacting bandwidth of the central moir\'e bands~\cite{Huder2018Electronic,Mesple2021Heterostrain}. At charge neutrality, strain has been argued to drive a transition from a correlated insulator to a semimetal~\cite{Parker2021Strain,hofmann2026monte,Liu2021Nematic}, the latter being consistent with
most experiments~\cite{xiaoImaging2026}. Furthermore, the incommensurate Kekulé spiral (IKS) state has been experimentally identified using scanning tunneling microscopy as the ground state for typical devices of TBG with heterostrain at filling factors near $\nu=\pm2$~\cite{nuckolls2023quantum,Kim2023imaging}, after predictions from Hartree-Fock (HF) mean field calculations of a gapped IKS at $\nu=\pm2,\pm3$~\cite{kwan2021kekule,Wagner2022Global}. 

The IKS involves intervalley coherence (IVC) between the microscopic graphene valleys, thereby breaking $U_v(1)$-valley symmetry. While various candidate strong-coupling insulators in the unstrained limit, such as the so-called Kramers intervalley coherent (KIVC) state~\cite{bultinck2020ground,lian2021TBG4}, also exhibit IVC, the IKS is unique in that the valleys only hybridize with an incommensurate moir\'e spiral wavevector $\bq_\text{IKS}$. This further breaks the moir\'e translation symmetry~\cite{kwan2021kekule}.

The IKS also differs from the well-studied strong-coupling insulators in other important ways. The strong-coupling states closely resemble simple quantum Hall ferromagnets, and are rooted in an emergent $U(4)\times U(4)$-symmetric limit of TBG which accommodates exact Slater determinantal ground states~\cite{bultinck2020ground,lian2021TBG4,kang2019strongcoupling,liaoCorrelation2021,ledwith2021strong}. On the other hand, the IKS is stabilized in the `intermediate-coupling' regime where interactions and kinetic dispersion compete on a similar footing, which has so far precluded exactly solvable limits and complicated analytical approaches~\cite{herzog2025thfstrain,herzog2025HFIKS}. At a mean-field level, it has also been shown that the Euler topology of the central bands~\cite{song2019all,ahn2019failure}, which is protected by combined $C_2$ and spinless time-reversal symmetry $\mathcal{T}$, imposes a complex momentum-space texture of the IVC in the IKS~\cite{kwan2025textured,Wagner2022Global}. However, it is not clear how the phase diagram and topological constraints of the IKS behave beyond mean-field theory.

The above experimental and theoretical considerations motivate quantum many-body numerical calculations of the phase diagram of strained TBG and the IKS. Previous density matrix renormalization group (DMRG) studies have investigated the collapse of the KIVC with strain at $\nu=$ -2~\cite{Parker2021Strain}, and the competition between the IKS and other phases at $\nu=$ -3~\cite{wang2023groundDMRG}. Here we perform approximated  continuous field momentum-space quantum Monte Carlo (AQMC) calculations~\cite{huangAngle2025,zhangMomentum2021}, exact diagonalization (ED), and HF simulations at $\nu=-2$. All three methods capture the relatively strong KIVC order at zero strain. At finite strains, both ED and HF capture IKS order, while AQMC fails to detect the relatively weak IKS order. 

\begin{figure}[t!]
\centering
\includegraphics[width=\linewidth]{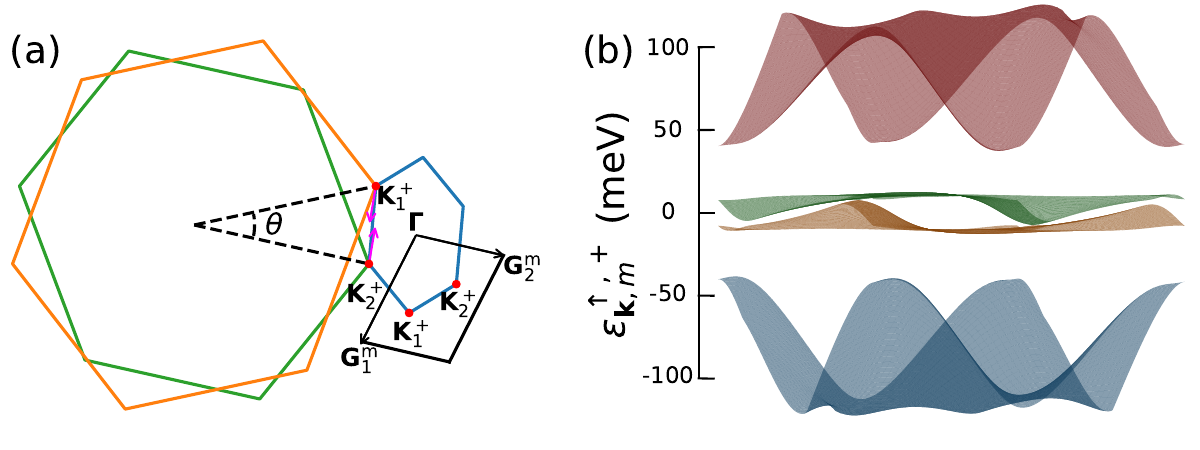}
\caption{Strained Brillouin zones and shifted Dirac cones. (a) Schematic Brillouin zones (green and orange hexagons) of two graphene layers with twist $\theta$ and strain of $\mp\epsilon_\s/2$ for each layer. The corresponding strained mBZ in the $\eta=+$ valley is the blue hexagon or black rhombus, with $\mathbf{\Gamma}$ the mBZ center, $\bG^\m_{1,2}$ the moir\'e RLV, and $\bK^+_{1,2}$ the mBZ corners. The magenta arrows show the exaggerated deviations of the monolayer Dirac cones. (b) The low-energy moir\'e bands with strain $\epsilon_\s = $ 0.3\% for valley $\eta=+$ and spin $s= \uparrow$.} 
\label{fig:BZ-Ek}
\end{figure}

\begin{figure*}[t!]
\centering
\includegraphics[width=\linewidth]{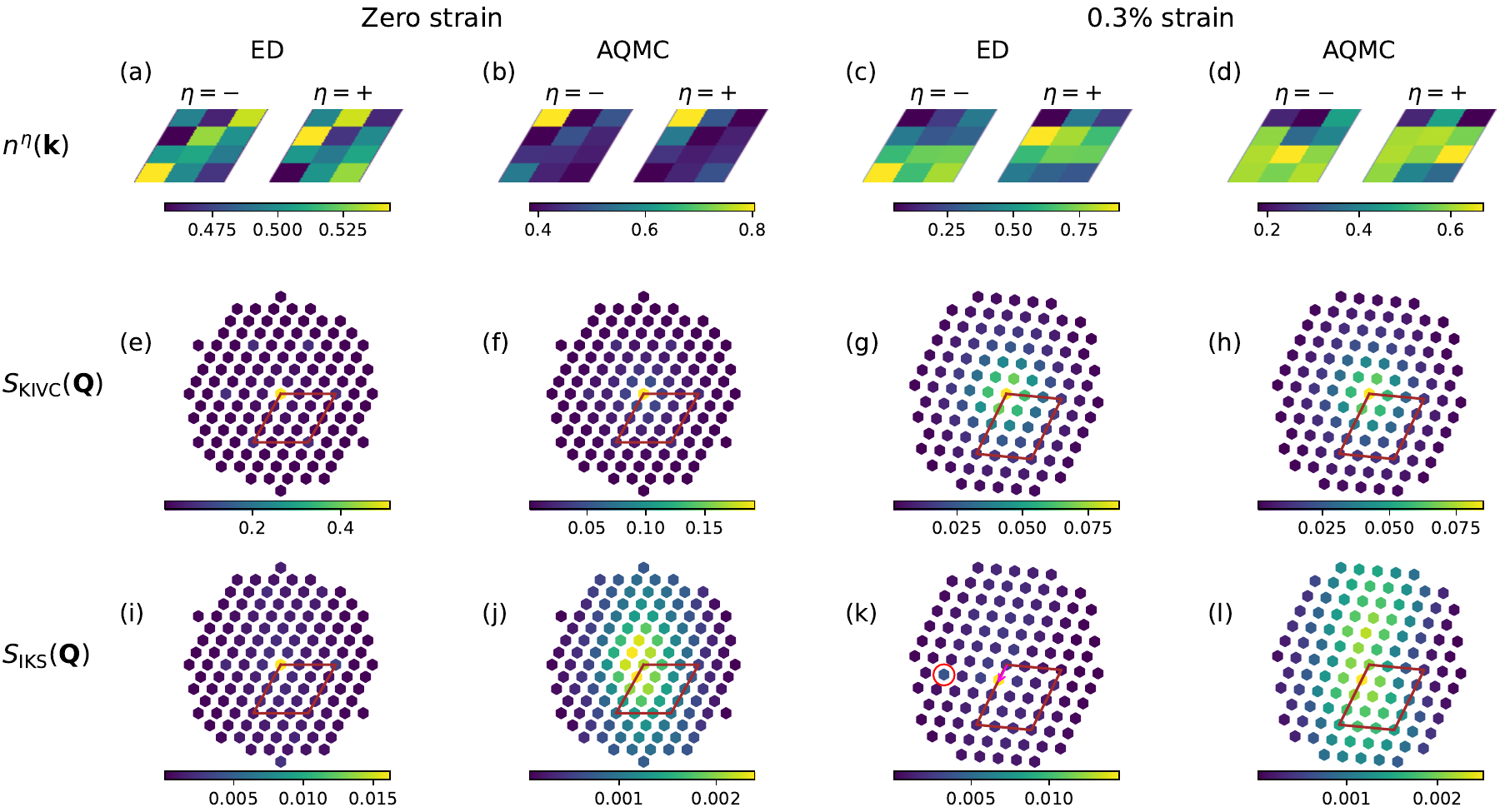}
    \caption{Occupation number ($n^\eta(\bk)$) and structure factors for KIVC ($S_\mathrm{KIVC}(\bQ)$) and IKS ($S_\mathrm{IKS}(\bQ)$) from exact diagonalization (ED) and approximated quantum Monte Carlo (AQMC). The system size is $4\times3$ with two strain strengths of $\epsilon_\s = $ 0.0\%, 0.3\%. We consider a spinless calculation at a spinless filling factor of $\nu=-1$, which is related to spinful $\nu=-2$. (a-d) show $n^\eta(\bk)$ for two valleys $\eta=\pm$ from ED and AQMC. (e-h) show $S_\mathrm{KIVC}(\bQ)$, where the mBZ is denoted as the brown rhombus, and (i-l) show $S_\mathrm{IKS}(\bQ)$. In (k), the magenta arrow indicates the finite $\bq_\mathrm{IKS}$ while the red circle indicates an additional Bragg peak. The color bar is shown below each plot.} 
\label{fig:nk-Sq-strn0-strn3}
\end{figure*}

\begin{figure*}[t]
\centering
\includegraphics[width=0.9\linewidth]{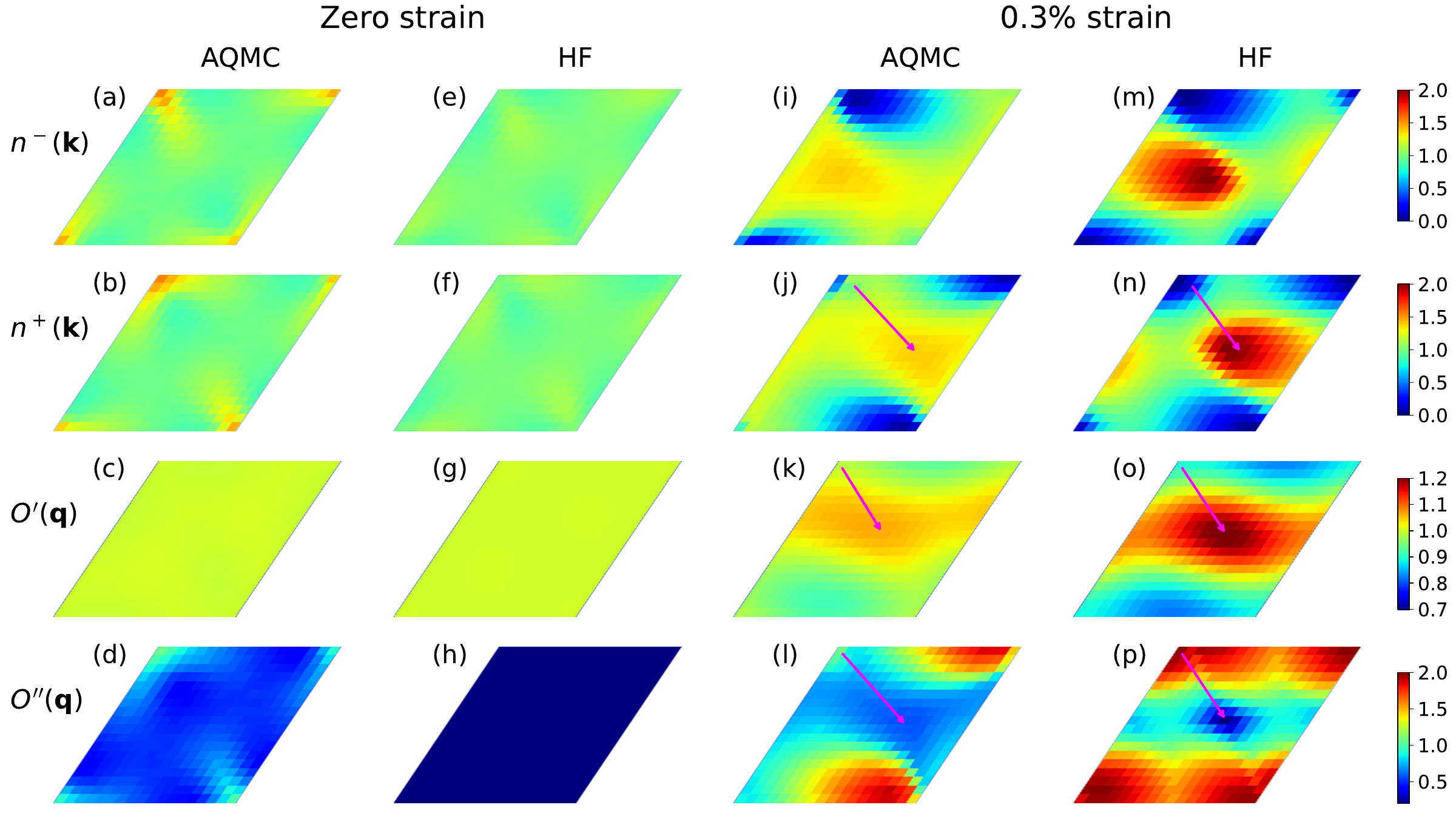}
\caption{Occupation $n^\eta(\bk)$ and corresponding nesting diagnostics from approximated quantum Monte Carlo (AQMC) and Hartree Fock (HF) at $\nu=-2$ with $N_{\bk}=18\times18$. (a)-(h) are with zero strain while (i)-(p) are with 0.3\% strain. (a)-(d) and (i)-(l) are from AQMC while (e)-(h) and (m)-(p) are from HF. The first two rows show the occupation factor $n^\eta(\bk)$ for valley $\eta=-$ and valley $\eta=+$ respectively. The magenta arrow in (j) points from the minimum in (i) to the maximum in (j), and similarly for the arrow in (n). The magenta arrow in (k) and (o) [(l) and (p)] corresponds to the maximum [minimum] of the IKS nesting diagnostic $O'(\bq)$ [$O''(\bq)$].
The HF ground state has IKS wavevector $\mathbf{q}_\text{IKS}=(8/18,9/18)$, which is the same as the magenta arrows in (n)-(p). Each row has its color bar at the rightmost column.
}
\label{fig:nk-Oq-strn0-strn3}
\end{figure*}


\section{Model and method}
For $\theta=1.15^\circ$ twisted bilayer graphene, we add uniaxial heterostrain $\epsilon_{\s,1}=-\epsilon_{\s,2}=-\frac{\epsilon_\s}{2}$ with the direction $\phi_{\s,1}=\phi_{\s,2}=\phi_\s=130^\circ$ for each layer (see Appendix~\ref{sec:interacting_continuum} for details of the strain implementation). As illustrated in Fig.~\ref{fig:BZ-Ek}(a), strain not only deforms the moir\'e reciprocal lattice vectors (RLVs), but also shifts the positions of the monolayer Dirac cones, both in momentum and energy~\cite{bi2019designing}. The kinetic Hamiltonian can be written as (see Appendix~\ref{sec:interacting_continuum})
\begin{equation}
    H_0=\sum_{s,\eta,\bk,m}\left(\epsilon^{s,\eta}_{\bk,m}-\mu\right)c^\dagger_{s,\eta,\bk,m} c_{s,\eta,\bk,m},
\end{equation}
where spin $s=\uparrow,\downarrow$, valley $\eta=\pm$, $\bk$ is the momentum in the moir\'e Brillouin zone (mBZ), $m$ is the band index and $\mu$ is the chemical potential. The  mBZ is discretized into a momentum grid with $N_{\bk}=L_1\times L_2$ points, corresponding to the number of moir\'e unit cells. Strain significantly affects the moir\'e band structure $\epsilon^{s,\eta}_{\bk,m}$ as shown in Fig.~\ref{fig:BZ-Ek} (b) for $\epsilon_\s=0.3\%$, and the evolution of two low-energy bands with $\epsilon_\s$ is shown in Appendix~\ref{sec:projection}.

Long-range density-density interactions are incorporated as
\begin{equation}
    H_\I=\sum_{\bQ>0}\frac{1}{4\Omega}V(\bQ)\left(A^2_\bQ-B^2_\bQ\right)
    \label{eq:HI_A_B}
\end{equation}
where $A_\bQ=\delta\rho_{-\bQ}+\delta\rho_{\bQ},B_\bQ=\delta\rho_{-\bQ}-\delta\rho_{\bQ}$, $\Omega$ is the total real-space system area, and the summation over momentum transfers $\bQ>0$ runs over half of momentum space (i.e.~only one of $\bQ$ or $-\bQ$ is included) and excludes $\bQ=0$. The density operator (see Appendix~\ref{sec:interacting_continuum} and \ref{sec:projection} for the derivation of $\delta\rho_{\bQ}$) is 
\begin{equation}
\begin{aligned}
\delta\rho_{\bQ}=&\sum_{s,\eta,k,m,n}\lambda^{s,\eta}_{m,n}\left(\bk,\bQ\right)\times\\
&\left(c^\dagger_{s,\eta,\bk,m}c_{s,\eta,\bk+\bQ,n}-\frac{1}{2}\delta_{\bQ\in\text{RLV}}\delta_{m,n}\right).
\label{eq:density_operator}
\end{aligned}
\end{equation}
The form factor 
$\lambda^{s,\eta}_{m,n}\left(\bk,\bQ\right)$
arises from the rotation from the plane-wave basis to the band basis. Since we have projected to the central bands, the $\frac{1}{2}$ term in Eq.~\eqref{eq:density_operator} corresponds to the `central average' interaction scheme utilized in e.g.~Refs.~\cite{bultinck2020ground,lian2021TBG4}, which preserves the approximate many-body particle-hole symmetry. The single-gate-screened Coulomb potential $V(\bQ) =\frac{e^2}{2 \varepsilon_0\varepsilon_\r|\bQ|}\left(1-\mathrm{e}^{-|\bQ| d}\right)$, with a screening parameter $d=20\,$nm and dielectric constant $\epsilon_\r=10$. $\bQ$ is cut off at the length of the longer primitive moir\'e RLV, due to the decay of the Coulomb potential and form factors~\cite{huangAngle2025}. 

We tackle the total Hamiltonian $H=H_0+H_\I$ at the filling of $\nu=-2$ using ED, AQMC, and HF (see Appendices~\ref{sec:ed}-\ref{sec:AQMC} for details). To bypass the exponential sign problem in AQMC~\cite{panSign2024}, we make an approximation by taking the absolute value of the original sampling weight as the new sampling weight. We show that such an approximation preserves moir\'e translation, $C_2\mathcal{T}$, and valley-$U(1)$ symmetries, which are relevant symmetries for the many-body phases we are interested in (see Appendix~\ref{sec:pAQMC} for details, which includes Figs. \ref{fig:Gt0_L6_strn6}, \ref{fig:Sqt_L6_strn6_KIVC}, and \ref{fig:Sqt_L6_strn6_IKS}). The consistency between stringent quantum Monte Carlo (QMC) and AQMC is also shown in Fig.~\ref{fig:QMC-AQMC} in Appendix~\ref{sec:Benchmark_ED_AQMC} at a high temperature.

To compare and physically understand the outputs of these techniques, we compute the occupation numbers and the intervalley structure factors. The former is a one-body correlation function 
\begin{equation}
    n^{\eta}(\bk)=\langle\sum_{s,m}c^\dagger_{s,\eta,\bk,m}c_{s,\eta,\bk,m}\rangle,
\end{equation}
which takes values between 0 and 4. Note that for a spinless calculation, the spin index $s$ is neglected, and $n^\eta(\bk)$ can only take values between 0 and 2. 

The formation of long-range IVC order can be diagnosed by the two-body intervalley structure factor 
\begin{gather}\label{eq:def_SQ}
S_{ij}(\bQ)=\frac{1}{N_{\bk}^2}\langle \hat{O}^\dagger_{ij}(\bQ)\hat{O}_{ij}(\bQ)\rangle,
\end{gather}
with the intervalley operator in the plane wave basis 
\begin{equation}
    \hat{O}_{ij}(\bQ)=\sum_{\bk\in\text{all}}\sum_{s l l' \sigma\sigma'}\hat{{c}}_{s,+,l,\sigma}^\dagger(\bk+\bQ)[\mu_{i}]_{ll'}[\sigma_{j}]_{\sigma\sigma'}\hat{{c}}_{s,-,l',\sigma'}(\bk). 
    \end{equation}
    $\mu_i$ and $\sigma_j$ are Pauli matrices in layer ($l$) and sublattice ($\sigma$) spaces, respectively. The KIVC symmetry-breaking order is detected in the $(\mu_2,\sigma_1)$ channel and invisible in the $(\mu_0,\sigma_0)$ channel due to its $\mathcal{T}'=\eta_y\mathcal{T}$ symmetry, while the IKS order is detected by $(\mu_0,\sigma_0)$ channel and invisible in the $(\mu_2,\sigma_1)$ channel due to its $\mathcal{T}$ symmetry. Hereafter $S_{2,1}$ and $S_{0,0}$ will be denoted as $S_\mathrm{KIVC}$ and $S_\mathrm{IKS}$ respectively. A non-vanishing value of the structure factor in the thermodynamic limit at certain wavevectors indicates the presence of the corresponding order. An IKS with spiral wavevector $\bq_{\text{IKS}}$ is expected to have `Bragg' peaks in $S_\mathrm{IKS}(\bq_{\text{IKS}}+\bG)$. As $N_\bk\rightarrow\infty$, the structure factor will be dominated by these Bragg peaks with finite amplitude, while other diffuse background features should decay to zero.

\section{Results}

We first compare and contrast spinless calculations between AQMC and ED on a system of size $4\times 3$. We focus on spinless $\nu=-1$, which is related to spinful $\nu=-2$ with a spin polarized ground state. We note that for the ED
calculation, the $4\times 3$ calculation contains 48 single-particle
orbitals (12 mBZ momenta with 2 bands and 2 valleys). The
largest symmetry sector has a Hilbert space dimension of $\sim 1.5\times 10^9$, which is of the same order or larger than those
considered in previous ED studies in TBG~\cite{2021TBG6,potasz2021ED}. 

The upper rows of Fig.~\ref{fig:nk-Sq-strn0-strn3} 
compare $n^{\eta}(\bk)$ computed using the two methods at zero strain and $\epsilon_\s = 0.3\%$. At zero strain, the patterns are qualitatively similar except for the $\Gamma$ point ($\bk=0$), where $n^\eta(\bk)$ in AQMC is enhanced. At $\epsilon_\s = 0.3\%$, ED shows significantly  inhomogeneous electronic occupation across the mBZ, with minima at $\Gamma$. The $n^\eta(\bk)$ in AQMC also become more inhomogeneous, similarly exhibiting a single pronounced minimum and maximum per valley. At the same time, the occupation fluctuation is suppressed compared to ED, and the peaks occur at different $\bk$. Note that the occupation distribution from AQMC is qualitatively consistent to that of HF for a much larger system, as we demonstrate later in Fig.~\ref{fig:nk-Oq-strn0-strn3}. 

The KIVC and IKS structure factors are shown in the two lower rows of Fig.~\ref{fig:nk-Sq-strn0-strn3}. Both ED and AQMC capture the KIVC order at $\Gamma$ at zero strain as shown in (e) and (f). At the strain of $\epsilon_\s = 0.3\%$, the KIVC order disappears in (g) and (h), as inferred by the broadening and reduction of the $S_{\text{KIVC}}$ peak compared to the situation at zero strain. Interestingly, ED resolves finite-$\bq$ peaks in $S_{\text{IKS}}$ [see panel (k)] with the most intense peak indicated by the magenta arrow and an additional Bragg peak indicated by the red circle. The location of these peaks are consistent with the estimation of $\bq_\text{IKS}$ extracted from the  maxima and minima of $n^\eta(\bk)$, according to the `lobe principle' explained later. These results are suggestive that the ED calculation resolves an IKS state. However, AQMC fails to generate such a sharp Bragg peak in $S_{\text{IKS}}$. Instead, we only observe an enhanced but broadened maximum at the same finite $\bq$ as shown in (l). Note that in ED the intensity of the IKS peak is much weaker than that of the KIVC peak. The relative weakness of the IVC amplitude in the IKS can be motivated by the fact that it arises in the intermediate-coupling regime, and the topological requirement of a momentum-space valley texture at the mean-field level~\cite{kwan2025textured}.  A possible interpretation of our results is that AQMC is unable to detect ordering with such weak magnitude.

In Fig.~\ref{fig:QMC-AQMC-ED-L2} in Appendix~\ref{sec:Benchmark_ED_AQMC}, we also provide comparisons of both $n^{\eta}(\bk)$ and the intervalley structure factors at $\epsilon_\s = 0.6\%$ for the $2\times2$ system, where  numerically exact QMC can be carried out. 
Moving forward, based on the above results, we take as a hypothesis that the AQMC adequately captures the qualitative features of $n^{\eta}(\bk)$ and $S_\mathrm{KIVC}(\bQ)$ for larger systems.

For a much larger system of $18\times18$, Fig.~\ref{fig:nk-Oq-strn0-strn3} shows $n^{\eta}(\bk)$ obtained from both AQMC and HF for both zero strain and $\epsilon_\s=0.3\%$ strain respectively. For zero strain, the occupation distribution is very homogeneous in HF and mainly hovers around 1, as shown in Fig.~\ref{fig:nk-Oq-strn0-strn3} (e) and (f).  While the occupation number in AQMC is mostly homogeneous, it still yields pronounced values around $\Gamma$, as shown in (a) and (b). The ground state is a (spin-polarized) KIVC state indicated by the KIVC structure factor in Fig.~\ref{fig:Sq-QMC-HF-strn0-strn6} (a) and (e) of Appendix~\ref{sec:SI_IVC}. At $\epsilon_\s=0.3\%$, qualitative agreement is found in $n^\eta(\bk)$ from AQMC and HF, with the occupation number distribution highly inhomogeneous across the mBZ in Fig.~\ref{fig:nk-Oq-strn0-strn3} (i), (j), (m) and (n). There is a strong depletion of electron population around $\Gamma$, while there is a `lobe' of high electron occupation at a non-zero $\bk$. We note that in the HF framework, $n^\eta(\bk)$ would be integer-valued in the absence of IVC. Hence, the fact that $n^\eta(\bk)$ evolves continuously with $\bk$ in panels (m) and (n) can be traced directly to the presence of IVC. In terms of quantitative differences between the AQMC and HF, the occupation numbers in the AQMC (panels (i) and (j)) do not reach as large as a value compared to HF where $n^\eta(\bk^\eta_\text{max})=2$. We attribute this observation to the fact that the AQMC contains more fluctuation effects than HF.

To connect the above observations to IKS physics, we first note that a mean-field IKS exactly satisfies $n^+(\bk+\bq_\text{IKS})+n^-(\bk)=\bar{n}$, where $\bar{n}=2$ is the average occupation factor (including both spins and valleys) over the mBZ. It has further been theoretically shown that  $n^\eta(\bk)$ of the IKS, which preserves $C_2$ and $\mathcal{T}$ symmetry, cannot be homogeneous over the mBZ, at least at mean-field level~\cite{kwan2025textured,wang2025CTI}. This arises from the non-trivial $C_2\mathcal{T}$-protected Euler index of the pair of central bands within each spin-valley flavor~\cite{song2019all,ahn2019failure}, which topologically mandates them to be connected by Dirac points. Any single $C_2\mathcal{T}$-symmetric band constructed from this pair is hence necessarily singular. In the projected model, the insulating IKS at $\nu=-2$ however contains only a single occupied IVC band within each spin sector. This forces $n^\eta(\bk)$ to take the minimal ($n^\eta(\bk)=0$) and maximal values ($n^\eta(\bk)=2$) somewhere in the mBZ~\cite{kwan2025textured} in order to `hide' these singularities and form a well-behaved insulator. Therefore, unlike a strong-coupling IVC state such as the KIVC, the valley-resolved occupation factor of a mean-field IKS is topologically required to modulate significantly across the mBZ.

The above considerations enable an estimation of the wavevector $\bq_\text{IKS}$ if an IKS state is indeed present. The idea is to match the location $\bk^-_\mathrm{min}$ of the occupation factor minimum in valley $\eta=-$ to the location $\bk^{+}_\mathrm{max}$ of the maximum in valley $\eta=+$. In the mean-field IKS, these momenta will hybridize with each other to satisfy $n^+(\bk+\bq_\text{IKS})+n^-(\bk)=\bar{n}$. From this, we therefore extract $\bq_\text{IKS}=\bk^{+}_\mathrm{max}-\bk^-_\mathrm{min}$, which is plotted as the magenta arrows in Fig.~\ref{fig:nk-Oq-strn0-strn3} (j) and (n).  

To assess the relevance of IKS physics in greater detail, we introduce two additional `nesting' diagnostics for IKS physics based on $n^\eta(\bk)$:
\begin{equation}
\begin{aligned}
O'(\bq)= &\frac{1}{N_\bk}\sum_{\bk}n^+(\bk+\bq)\left(2-n^{-}(\bk)\right),\\
O''(\bq)= &\mathrm{max}_\bk|n^+(\bk+\bq)+n^-(\bk)-2|\geq0.\label{eq:O''q}
\end{aligned}
\end{equation}
For a mean-field IKS, $O''(\bq_\text{IKS})$, which has been introduced previously in the DMRG study of Ref.~\cite{wang2023groundDMRG}, vanishes identically at the IKS wavevector. Hence, more generally, the value of $\bq$ at which $O''(\bq)$ reaches its minimum provides an estimate for the favored IVC wavevector $\bq_\text{IKS}$. Similarly, for a mean-field IKS, $O'(\bq)$ is expected to be maximal at $\bq_\text{IKS}$, since this is when the maximal occupation in one valley coincides with the minimal occupation in the other valley, and vice versa. 
The position of the maximum (minimum) of $O'(\bq)$ ($O''(\bq)$) therefore provides an estimate of the most favorable wavevector for IVC in the IKS ground state. Even if long-range IVC order is not strictly present, these nesting diagnostics can be viewed as predictions of the wavevector of a putative incipient IKS.


\begin{figure}[t]
\centering
\includegraphics[width=\linewidth]{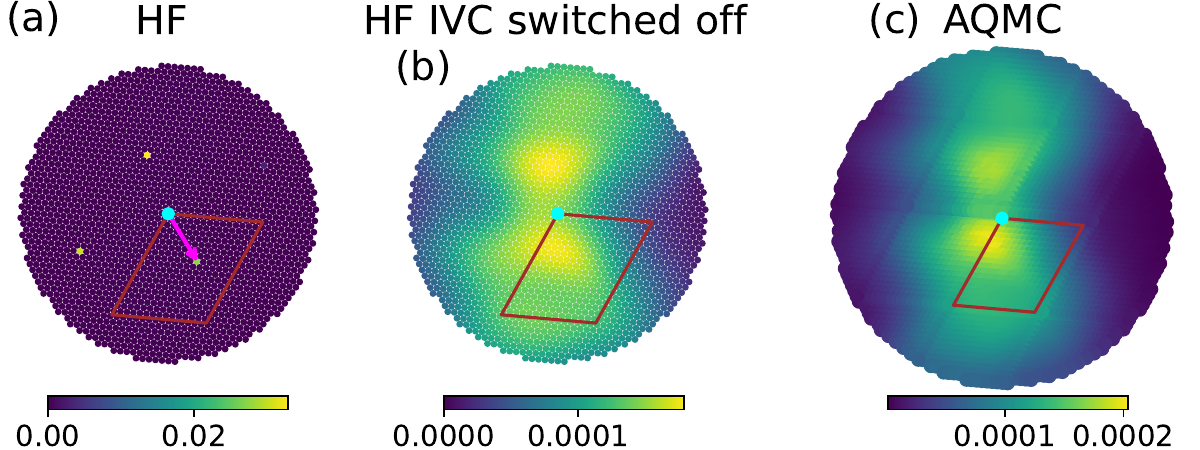}
\caption{Intervalley structure factor $S_\mathrm{IKS}(\bQ)$ from Hartree Fock (HF) and approximated quantum Monte Carlo (AQMC) at $\nu=-2$ with $N_{\bk}=18\times18$ and $\epsilon_\mathrm{s} = $ 0.3\%. (a) $S_\mathrm{IKS}(\bQ)$ from HF with Bragg peaks corresponding to $\bq_\text{IKS}=(8/18,9/18)$. The magenta arrow points to the dominant peak and is the same as those in Fig.~\ref{fig:nk-Oq-strn0-strn3} (n)-(p). The brown rhombus is the mBZ and the cyan dot indicates $\bQ=0$. In (b), the IVC in the final HF density matrix has been artifically switched off when computing $S_\mathrm{IKS}(\bQ)$. (c) $S_\mathrm{IKS}(\bQ)$ from AQMC.}
\label{fig:S-HF-QMC-strn3}
\end{figure}

For zero strain, $O'(\bq)$ and $O''(\bq)$ do not reveal any noticeable features as shown in Fig.~\ref{fig:nk-Oq-strn0-strn3} (c), (d), (g), and (h). For $\epsilon_s=0.3\%$, the maximum of $O'(\bq)$ and the minimum of $O''(\bq)$ become prominent, and they look qualitatively similar between AQMC (Fig.~\ref{fig:nk-Oq-strn0-strn3} (k) and (l)) and HF (Fig.~\ref{fig:nk-Oq-strn0-strn3} (o) and (p)). The deviation of $\bq_\text{IKS}$ from these diagnostics compared to the simpler estimation $\bq_\text{IKS}=\bk^{+}_\mathrm{max}-\bk^-_\mathrm{min}$ is small, being within $<20\%$ of the mBZ size. As expected, $O''(\bq)$ has a minimum of zero in the HF calculation, while that in AQMC does not reach zero, but displays a significantly broadened minimum. This may be connected to the fact that the IKS wavevector is known to be soft in mean-field calculations~\cite{kwan2021kekule}, with closely-competing energy states as shown explicitly in Fig.~\ref{fig:allq_energy} of Appendix~\ref{sec:SI_Hartree_Fock_allq}.

For $\epsilon_\s=0.3\%$, we show the intervalley structure factor from HF [Fig.~\ref{fig:S-HF-QMC-strn3} (a)] and AQMC [Fig.~\ref{fig:S-HF-QMC-strn3} (c)]. As expected, the HF result exhibits peaks when $\bQ$ is equal to $\bq_\text{IKS}$ modulo a moir\'e RLV. However, AQMC fails to yield the corresponding IKS peaks as shown in Fig.~\ref{fig:S-HF-QMC-strn3} (c).
We comment that in HF, the intensity of the $S_{\text{IKS}}(\bQ)$ peak in the IKS state at finite strain is roughly one order of magnitude weaker than the $S_\text{KIVC}(\bQ)$ peak in the KIVC state at vanishing strain. This points to the strength of intervalley coherence in the IKS being relatively weaker, which can be traced to the topologically-enforced momentum-space modulation of the order parameter as discussed previously. This may explain why the AQMC does not reveal explicit signatures of IKS ordering in the structure factor.
In Fig.~\ref{fig:S-HF-QMC-strn3} (b), we isolate the diffuse background from Fig.~\ref{fig:S-HF-QMC-strn3} (a) by artificially switching off the IVC part of the self-consistent HF density matrix when computing $S_\mathrm{IKS}(\bQ)$. This procedure captures the effect of the IKS order on the renormalization of the flat bands, but effectively removes the long-range coherence.
Interestingly, we find a close agreement between AQMC (Fig.~\ref{fig:S-HF-QMC-strn3} (c)) and the diffuse background from HF (Fig.~\ref{fig:S-HF-QMC-strn3} (b)).

\section{Discussion}
While the AQMC structure factor does not exhibit intervalley Bragg peaks, the resemblance of the diffuse background of $S_{\text{IKS}}(\bQ)$ and the occupation number distribution with the mean-field IKS suggests that the AQMC captures some key aspects of the physics related to the IKS phase. 
For instance, one possibility is that short-range correlations and interaction effects in  AQMC have `primed' the electrons to form intervalley electron-hole pairs, but long-range coherence is not yet established due to thermal/quantum fluctuations. Mean-field theory is expected to overestimate symmetry-breaking order, especially for phases that are not connected to an exact mean-field limit, and it could be that the true many-body ground state does not have proper IKS order for the parameters chosen here. Even within HF, the IKS manifold consists of multiple nearly-degenerate solutions with different $\bq_\text{IKS}$ (see Fig.~\ref{fig:allq_energy} in Appendix~\ref{sec:SI_Hartree_Fock_allq}), and this softness of the ordering wavevector could factor into the absence of the IKS peak in AQMC. We comment that compared to the strong-coupling IVC states, the IKS in mean-field theory has a comparatively weaker total magnitude of IVC owing to its topologically-enforced momentum-space modulation.
Another possibility is that AQMC is fundamentally unable to capture relatively weak peaks in the structure factor for the IKS. 

Nevertheless, the single-particle Green's function in AQMC decays exponentially as a function of imaginary time for all $\bk$ (Fig.~\ref{fig:Gt0} in Appendix~\ref{sec:SIII}), implying a gapped insulating ground state from the electronic perspective. If the first scenario outlined above is applicable, then our results point towards a valley-disordered but charge-gapped state. In the second scenario, the fact that both AQMC and HF yield electronically-insulating states further suggests the interpretation of the IKS as consisting of a few auxiliary field configurations (a few Slater determinant
states), whose one-particle properties can be captured by AQMC, while it misses the two-particle correlator, i.e.~$S_{\text{IKS}}(\bQ)$.

Our results point to the `intermediate-coupling' and the topologically constrained nature of the IKS state in a projected correlated flat-band setting. As the AQMC can still capture a strong order like KIVC, as further shown in Fig.~\ref{fig:Sq-QMC-HF-strn0-strn6} (a) of Appendix~\ref{sec:SI_IVC} for a much larger system size of 18$\times$18, our combined numerical protocol may provide new perspectives towards understanding the many-body ground state of TBG. An interesting future direction is to examine the usefulness of these techniques to non-integer fillings where HF has even less {\it a priori} justification for adequately capturing the relevant physics. It would be also be worthwhile to assess the applicability of AQMC to examine finite-temperature physics~\cite{zhangSuperconductivity2022,panThermodynamic2023}, such as the thermal disordering of the correlated states and energy scales of different elementary excitations in the system~\cite{linExciton2022,panThermodynamic2023,ledwithNonlocal2025,ledwithExotic2025,shanTrion2026}. 

\section{Acknowledgments}
We acknowledge discussions with Nikolaos Parthenios, Jeyong Park and Patrick Ledwith on similar topics. C.H. and Z.Y.M. acknowledge the support from the Research Grants Council (RGC) of Hong Kong Special Administrative Region (SAR) of China (Project Nos. AoE/P701/20, C7037-22GF, 17302223, 17301924,
17301725, AoE/P604/25R), the ANR/RGC Joint Research Scheme sponsored by RGC of Hong Kong and French National Research Agency (Project No. A\_HKU703/22) and the State Key Laboratory of Optical Quantum Materials at HKU. C.H. and Z.Y.M thank HPC2021 system under the Information Technology Services at the University of Hong Kong~\cite{hpc2021}, as well as the Beijing Paratera Tech Corp., Ltd~\cite{paratera} for providing HPC resources that have contributed to the research results reported within this paper. L.C. was funded by the Simons Foundation SFI-MPS-NFS-00006741-11, the Deutsche Forschungsgemeinschaft (DFG, German Research Foundation) under Germany's Excellence Strategy-EXC-2111-390814868 and TRR 360-492547816, and by the European Union (ERC-2023-STG, Project 101115758 - QuantEmerge).
MU acknowledges support by  the Deutsche Forschungsgemeinschaft  through AS 120/19-1 (Project No.~530989922).
FFA acknowledges support by the W\"urzburg-Dresden Cluster of Excellence {\it ctd.qmat} (EXC 2147, Project No.~390858490). 



\appendix

\section{Interacting continuum model of strained TBG}\label{sec:interacting_continuum}
In graphene, uniaxial strain with strength $\epsilon_\s$ and orientation $\phi_\s$ (counter-clockwise) is characterized by the tensor
\begin{equation}
    S(\epsilon_\s,\phi_\s)=\begin{pmatrix}
\cos\phi_\s & \sin\phi_\s\\
-\sin\phi_\s & \cos\phi_\s
\end{pmatrix}\begin{pmatrix}
-\epsilon_\s & 0\\
0 & \nu_s\epsilon_\s
\end{pmatrix}\begin{pmatrix}
\cos\phi_\s & -\sin\phi_\s\\
\sin\phi_\s & \cos\phi_\s
\end{pmatrix},
\end{equation}
with Poisson ratio $\nu_\s=0.16$. We represent the combined action of counter-clockwise twist $R(\theta)$ with angle $\theta$ and uniaxial straining on a single layer as~\cite{bi2019designing} 
\begin{equation}
\mathbb{T}(\theta,\epsilon_\s,\phi_\s)=\left(R(\theta)+S(\epsilon_\s,\phi_\s)\right)^\mathrm{T}.
\end{equation}
Let $\bG^\g_{1,2}$ denote the two unstrained and non-rotated RLVs of graphene. Then, the moir\'e RLVs for TBG is
\begin{equation}
\bG^\mathrm{m}_{1,2}=\left(\mathbb{T}(\frac{\theta}{2},-\frac{\epsilon_\s}{2},\phi_\s)-\mathbb{T}(-\frac{\theta}{2},\frac{\epsilon_\s}{2},\phi_\s)\right)\bG^\mathrm{g}_{1,2},
\end{equation}
where $\mathbb{T}(\frac{\theta}{2},-\frac{\epsilon_\s}{2},\phi_\s)$ and $\mathbb{T}(-\frac{\theta}{2},\frac{\epsilon_\s}{2},\phi_\s)$ represent the rotation and straining of the upper ($l=1$) and lower ($l=2$) layers respectively. 
Similarly, the graphene BZ corners are rescaled as 
\begin{equation}
\bK^\eta_l=\bbT\left((-1)^{l+1}\frac{\theta}{2},(-1)^{l}\frac{\epsilon_\s}{2},\phi_\s\right)\left(\frac{|\bG_1^\g|}{\sqrt{3}},0\right)^\T
\end{equation}
for each layer.

Hereafter, we use $\bG_{1,2}$ to refer to the moir\'e reciprocal lattice vectors $\bG^\mathrm{m}_{1,2}$. The single-particle BM continuum Hamiltonian~\cite{bistritzerMoire2011} can be constructed in the plane-wave basis as
\begin{equation}
\begin{aligned}
H_0=&\sum_{s,\eta,\bk,\bG,X,\bG',X'}c^\dagger_{s,\eta,X}(\bk+\bG)\left(H^{s,\eta}_\bk\right)_{\bG,X,\bG',X'}
\\
&\times c_{s,\eta,X}(\bk+\bG')\\
=&\sum_{s,\eta,\bk}\vc^\dagger_{s,\eta}(\bk)H^{s,\eta}_\bk
\vc_{s,\eta}(\bk).
\end{aligned}
\label{eq:eq0}
\end{equation}
Here $c^\dagger_{s,\eta,X}(\bk+\bG)$ is the plane wave creation operator for spin $s$, valley $\eta$, and layer-sublattice index $X\in\{1\mathrm{A}, 1\mathrm{B}, 2\mathrm{A}, 2\mathrm{B}\}$.
The plane wave momentum $\bk+\bG$ is decomposed into a part $\bk$ lying within the mBZ, and the remainder $\bG$ which lies on the moir\'e reciprocal lattice.
$\vc^\dagger_{s,\eta}(\bk)$ is a vector that collects $c^\dagger_{s,\eta,X}(\bk+\bG)$ for all values of $\bG$ and $X$. Therefore, $H^{s,\eta}_\bk$ is a $4N_\bG\times 4N_\bG$ matrix, where $N_\bG$ is the number of plane waves retained within a momentum cutoff, $|\bG|\leq6\times\mathrm{max}\{|\mathbf{G}_1|,|\mathbf{G}_2|\}$. The matrix elements of $H^{s,\eta}_\bk$ are
\begin{equation}
	\begin{aligned}
		\left(H^{s,\eta}_\bk\right)_{\bG,\bG'}
        =\delta_{\bG,\bG'}\hbar v_\mathrm{F}
\left(\begin{array}{cc} 
			M^\eta_1  &  0  \\
			0  & M^\eta_2
		\end{array}\right) 
		+\left(\begin{array}{cc} 
			0  &  T^\eta_1  \\ 
			{T^\eta_2}^\dagger & 0
		\end{array}\right)
	\end{aligned}
	\label{eq:BM}.
\end{equation}
The first term corresponds to the intralayer Dirac term where
\begin{equation}
M^\eta_l=-(\bk+\bG-\bA_l^\eta-\bK_l^\eta) \cdot \pmb{\sigma}^\eta,
\end{equation}
and $\pmb{\sigma}^\eta = \left(\eta\sigma_x,\sigma_y\right)$ with $\sigma_x,\sigma_y$ Pauli matrices in sublattice space. The Dirac velocity $\frac{\hbar v_\mathrm{F}}{ \sqrt{3}a}=2.37745$ eV, where $a=1.42$ $\mathring{\text{A}}$ is the nearest neighbor carbon distance. The vector potential
\begin{equation}
\bA^\eta_l(\frac{\epsilon_\s}{2})=\eta\frac{3.14}{2a}\left({S(\frac{\epsilon^l_\s}{2})}_{11}-{S(\frac{\epsilon^l_\s}{2})}_{22},-2{S(\frac{\epsilon^l_\s}{2})}_{12}\right),
\end{equation}
captures the shift of the monolayer Dirac cones from the rotated and strained graphene BZ corners $\bK_l^\eta$. The second term of $H^{s,\eta}_\bk$ encodes the interlayer tunneling processes according to
\begin{equation}
\begin{aligned}
T^\eta_l=&\left(\begin{array}{cc}
u_0 & u_1\\
u_1 & u_0\end{array}\right)\delta_{\bG,\bG'}
+\left(\begin{array}{cc}
u_0 & u_1\mathrm{e}^{-i\frac{2\pi}{3}\eta}\\
u_1\mathrm{e}^{i\frac{2\pi}{3}\eta} & u_0\end{array}\right)\delta_{\bG,\bG'+(-1)^l\eta\G_1}\\
&+\left(\begin{array}{cc}
u_0 & u_1\mathrm{e}^{i\frac{2\pi}{3}\eta}\\
u_1\mathrm{e}^{-i\frac{2\pi}{3}\eta} & u_0\end{array}\right)\delta_{\bG,\bG'+(-1)^l\eta\left(\G_1+\G_2\right)},
\end{aligned}
\end{equation}
where $u_0$ and $u_1$ are the same-sublattice and opposite-sublattice tunneling amplitudes respectively. We set $u_0 = 80$ meV and $u_1=110$ meV. 

The long-range density-density interaction expressed in momentum space is
\begin{equation}
H_\mathrm{I}=\frac{1}{2\Omega}\sum_{\bq\in\mathrm{mBZ},\bG}V(\bq+\bG)\delta\rho_{\bq+\bG}\delta\rho_{-\bq-\bG},
\end{equation}
where
\begin{equation}
\begin{aligned}
&\delta\rho_{\bq+\bG}\\
=&\sum_{s,\eta,\bk,\bG',X'}
c^\dagger_{s,\eta,X'}(\bk+\bG')c_{s,\eta,X'}(\bk+\bq+\bG+\bG')\\
&-\sum_{s,\eta,\bk,\bG',X'}
\frac{1}{2}\delta_{\bq,0}\delta_{\bG,0}\\
=&\sum_{s,\eta,\bk}
\left(\vc^\dagger_{s,\eta}(\bk)\vc_{s,\eta}(\bk+\bq+\bG)-\sum_{\bG',X'}\frac{1}{2}\delta_{\bq,0}\delta_{\bG,0}\right)
\end{aligned}
\end{equation}
is the electron density operator measured relative to a reference point that is charge neutral and featureless. 
$\Omega$ is the total real-space area of the system. 

\section{Projection from plane wave basis to band basis}
\label{sec:projection}
\begin{figure*}[!ht]
\centering
\includegraphics[width=0.9\linewidth]{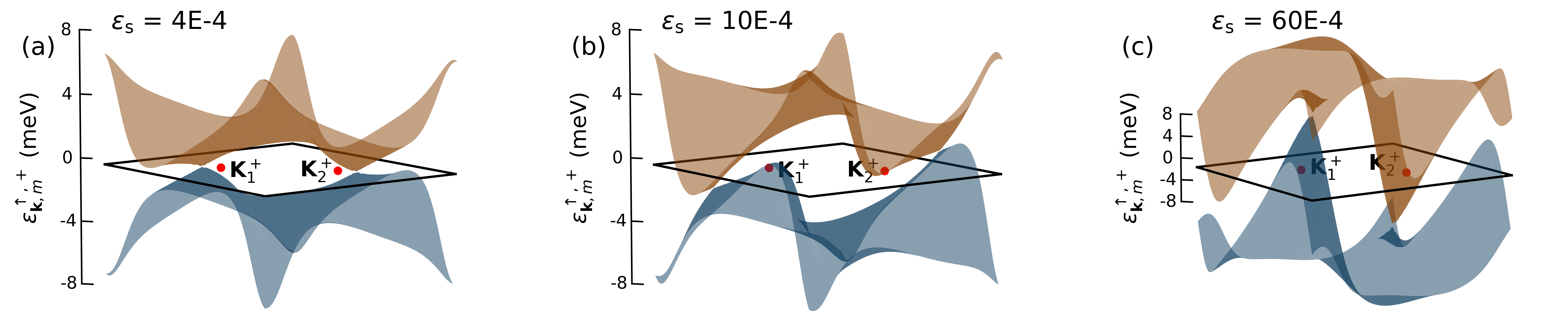}
\caption{(a)-(c) Evolution of the non-interacting dispersion with the strain strength $\epsilon_\s$. Note that $\theta=1.15^\circ,\phi_s=130^\circ, u_0=80\,$meV, and the moir\'e Brillouin zone is denoted by the black rhombus. Red dots stand for the moir\'e Brillouin zone corners $\bK^+_1$ and $\bK^+_2$. Different dispersion colors are used to indicate distinct bands. 
}
\label{fig:band1}
\end{figure*}

The momentum transfer $\bq+\bG$ is denoted as $\mQ$ in the following. We denote the eigenvalues and eigenvectors of the single-particle continuum model $H_0$ as $\epsilon^{s,\eta}_{\bk,m}$ and $\left|u^{s,\eta}_{\bk,m}\right>$, satisfying $H_0\left|u^{s,\eta}_{\bk,m}\right>=\epsilon^{s,\eta}_{\bk,m}\left|u^{s,\eta}_{\bk,m}\right>$, where $m$ is a moir\'e band index. This can also be represented as the matrix equation
\begin{equation}
H^{s,\eta}_\bk U^{s,\eta}_\bk=U^{s,\eta}_\bk E^{s,\eta}_\bk,
\end{equation}
where the $m$'th column of $U^{s,\eta}_\bk$ consists of the eigenvector for band $m$, and $E^{s,\eta}_\bk$ is a diagonal matrix whose entries consist of the $\epsilon^{s,\eta}_{\bk,m}$. Band structures computed with varying strains are shown in Fig.~\ref{fig:band1},  where the inflation of the bandwidth of the two low-energy bands and the shifting of the deformed Dirac cones from the mBZ corners are seen. 

To express $H_0$ in the band basis, we first write
\begin{equation}
\begin{aligned}
H_0=&\sum_{s,\eta,\bk}\vc^\dagger_{s,\eta}(\bk)H^{s,\eta}_\bk
\vc_{s,\eta}(\bk)\\
=&\sum_{s,\eta,\bk}\vc^\dagger_{s,\eta}(\bk) U^{s,\eta}_\bk E^{s,\eta}_\bk {U^{s,\eta}_\bk}^\dagger\vc_{s,\eta}(\bk).
\label{eq:proj_kinetics}
\end{aligned}
\end{equation}
We define the band basis operators
\begin{equation}
\vc_{s,\eta,\bk}^\dagger=\vc^\dagger_{s,\eta}(\bk) U^{s,\eta}_\bk,
\end{equation}
where the elements of $\vc_{s,\eta,\bk}$ are
\begin{equation}
c^\dagger_{s,\eta,\bk,m}=\sum_{\bG,X}U^{s,\eta}_{\bk,m,\bG,X}c^\dagger_{s,\eta,X}(\bk+\bG).\end{equation}
In this work, we work in the periodic gauge by imposing
\begin{equation}
[U^{s,\eta}_{\bk+\bG}]_{\bG',X}=[U^{s,\eta}_\bk]_{\bG'+\bG,X},
\end{equation}
such that $|u^{s,\eta}_{\bk+\bG,m}\rangle_{\bG',X}=|u^{s,\eta}_{\bk,m}\rangle_{\bG'+\bG,X}$ and $c_{s,\eta,\bk+\bG,m}=c_{s,\eta,\bk,m}$.
The single-particle continuum Hamiltonian is then recast as
\begin{equation}
\begin{aligned}
H_0=&\sum_{s,\eta,\bk}\vc_{s,\eta,\bk}^\dagger E^{s,\eta}_\bk \vc_{s,\eta,\bk}\\
=&\sum_{s,\eta,\bk,m}\epsilon^{s,\eta}_{\bk,m}c_{s,\eta,\bk,m}^\dagger c_{s,\eta,\bk,m}.
\label{eq:H0_methods}
\end{aligned}
\end{equation}
The projection to the two low-energy bands (per spin and valley) is performed by keeping only the values of $m$ corresponding to these bands.
We find it convenient to further constrain the gauge of the BM Bloch functions of the central bands. We choose the relative gauge between the valleys so that $\mathcal{PT}$ (which takes the form $\eta_x\mu_y\sigma_x$ in the plane wave basis) acts as $\eta_yn_y$ in the band basis with $n_y$ acting on bands.
In the ED calculations, we also fix the $C_2\mathcal{T}$ gauge (which acts in the plane wave basis according to $\sigma_x\mathcal{K}$, with $\mathcal{K}$ complex conjugation) to make the form factor real.

We now rewrite the interaction term. Using the same transformation, we find
\begin{equation}
\begin{aligned}
&\vc^\dagger_{s,\eta}(\bk)\vc_{s,\eta}(\bk+\mQ)\\
&=\sum_{m,n}c^\dagger_{s,\eta,\bk,m}\left({U^{s,\eta}_{\bk}}^\dagger U^{s,\eta}_{\bk+\mQ}\right)_{m,n}c_{s,\eta,\bk+\mQ,n}\\
&=\sum_{m,n}c^\dagger_{s,\eta,\bk,m}\lambda^{s,\eta}_{m,n}(\bk,\mQ)c_{s,\eta,\bk+\mQ,n},
\label{eq:proj_interaction}
\end{aligned}
\end{equation}
where the form factor is defined as
\begin{equation}
\begin{aligned}
\lambda^{s,\eta}_{m,n}(\bk,\mQ)=&\left({U^{s,\eta}_\bk}^\dagger U^{s,\eta}_{\bk+\mQ}\right)_{m,n}\\
=&\langle u^{s,\eta}_{\bk,m}|u^{s,\eta}_{\bk+\mQ,n}\rangle.
\end{aligned}
\end{equation}
This leads to
\begin{equation}
\delta\rho_\mQ=\sum_{s,\eta,\bk}\left(\sum_{m,n}\lambda^{s,\eta}_{m,n}(\bk,\mQ)c^\dagger_{s,\eta,\bk,m}c_{s,\eta,\bk+\mQ,n}-\sum_{\bG',X}\frac{1}{2}\delta_{\bq,0}\delta_{\bG,0}\right).
\end{equation}
Moreover, since 
\begin{equation}
\begin{aligned}
\sum_{\bG',X}\delta_{\bG,0}=&\sum_{m}\langle u^{s,\eta}_{\bk,m}|u^{s,\eta}_{\bk+\bG,m}\rangle\\
=&\sum_{m,n}\lambda^{s,\eta}_{m,n}(\bk,\bG)\delta_{m,n},
\end{aligned}
\end{equation}
and
\begin{equation}
\begin{aligned}
\sum_{\bG',X}\delta_{\bq,0}\delta_{\bG,0}=\sum_{m,n}\lambda^{s,\eta}_{m,n}(\bk,\mQ)\delta_{\bq,0}\delta_{m,n},
\end{aligned}
\end{equation}
we have
\begin{equation}\label{methodseq:rhoQ}
\delta\rho_\mQ=\sum_{s,\eta,\bk,m,n}\lambda^{s,\eta}_{m,n}(\bk,\mQ)\left(c^\dagger_{s,\eta,\bk,m}c_{s,\eta,\bk+\mQ,n}-\frac{1}{2}\delta_{\bq,0}\delta_{m,n}\right).
\end{equation}
Similar to the procedure for the single-particle Hamiltonian $H_0$, one method for implementing band projection for the interaction term $H_\I$ is to simply restrict the band indices to the central bands in $\delta \rho_{\mQ}$. This corresponds to the `central average' interaction scheme utilized in e.g.~Refs.~\cite{bultinck2020ground,lian2021TBG4}. We emphasize that the choice of interaction scheme is not unique (even when imposing constraints such as approximate many-body particle-hole symmetry). Within the projected Hamiltonian, different interaction schemes lead to different effective one-body corrections arising from the remote bands. Note that this interaction scheme differs from that used in Refs.~\cite{zhangFermion2022,zhangPolynomial2023}, where the last term in Eq.~\ref{methodseq:rhoQ} depended on the target filling factor $\nu$, which led to a polynomial instead of an exponential sign problem at  chiral non-zero integer fillings in the absence of strain.

Since the set of momentum transfers has inversion symmetry, we can therefore write
\begin{equation}
\begin{aligned}
H_\mathrm{I}&=\sum_{|\mathcal{Q}|\neq 0}\frac{1}{2\Omega}V(\mathcal{Q})\delta\rho_{\mathcal{Q}}\delta\rho_{-\mathcal{Q}}\\
&=\sum_{\bQ>0}\frac{1}{2\Omega}V(\bQ)(\delta\rho_{\bQ}\delta\rho_{-\bQ}+\delta\rho_{-\bQ}\delta\rho_{\bQ})\\
&=\sum_{\bQ>0}\frac{1}{4\Omega}V(\bQ)\left(\left(\delta\rho_{-\bQ}+\delta\rho_\bQ\right)^2-\left(\delta\rho_{-\bQ}-\delta\rho_\bQ\right)^2\right)\\
&=\sum_{\bQ>0}\frac{1}{4\Omega}V(\bQ)\left(A_\bQ^2-B_\bQ^2\right),
\end{aligned}
\label{eq:HI}
\end{equation}
where the summation over momentum transfers $\bQ>0$ runs over half of the set of $\{\mQ\}$ (i.e.~only one of $\bQ$ or $-\bQ$ is included). The $\bQ=0$ contribution can be absorbed into a tunable chemical potential term $\mu$. The total Hamiltonian is then
\begin{equation}
    H=H_0+H_\I-\mu\sum_{s,\eta,\bk,m}c_{s,\eta,\bk,m}^\dagger c_{s,\eta,\bk,m},
\end{equation}
where $\mu$ is used to adjust the average filling factor.

\section{Exact Diagonalization}
\label{sec:ed}
For the exact diagonalization calculations, we consider both valleys and restrict to a single spin. Since we employ product (Fock) states with definite particle number, the chemical potential term $\mu$ is not needed. The Hamiltonian can be expressed as
\begin{equation}
\begin{aligned}
H=&\sum_i\left(\epsilon_i+V_0(i)\right)c^\dagger_ic_i+\sum_{i\neq j}V_1(i,j)c^\dagger_ic_ic^\dagger_jc_j+H_1,
\end{aligned}
\end{equation}
where
\begin{equation}
\begin{aligned}
H_1=&\sum_{i>j}V_2(i,j)c^\dagger_ic_j+\sum_{i>j,k}V_3(i,j,k)c^\dagger_ic_jc^\dagger_kc_k\\&+\sum_{i>j>k>l}\left(V_4(i,j,k,l)c^\dagger_ic_jc_kc^\dagger_l+V_5(i,j,k,l)c^\dagger_ic_jc^\dagger_kc_l\right)\\&+\sum_{i>j>k>l}V_6(i,j,k,l)c^\dagger_ic^\dagger_jc_kc_l\\&+h.c..
\end{aligned}
\end{equation}

A product state is
\begin{equation}
\begin{aligned}
|\mathscr{i}\rangle=&c^\dagger_{i_1}c^\dagger_{i_2}\cdots c^\dagger_{i_N}|0\rangle\\
=&|0\eqnmark[black]{i_1}{1}0\ 0\eqnmark[black]{i_2}{1}\cdots\eqnmark[black]{i_N}{1}0\rangle,
\end{aligned}
\end{equation}
where $|0\rangle$ consists of $4N_\bk$ bits of 0 and $N_\bk$ corresponds to the particle number at filling $\nu=-1$ (i.e.~$\nu=-2$ in the spinful calculation). $i_i\in[1,4N_\bk]$ is a composite index for $c^\dagger_{\eta,\bk,m}$, and $i_1>i_2>\cdots>i_N$. In this basis, the diagonal elements of $H$ in the product space are
\begin{equation}
H_{\mathscr{ii}}=\sum_{|\mathscr{i}\rangle_i=1}\left(\epsilon_i+V_0(i)\right)+\sum_{|\mathscr{i}\rangle_i=|\mathscr{i}\rangle_j=1}V_1(i,j),
\end{equation}
where $|\mathscr{i}\rangle_i$ stands for the $i^\mathrm{th}$ element of $|\mathscr{i}\rangle$. The off-diagonal elements $H_{\mathscr{ij}}$ of the lower-triangular part of $H$ in the product space are
\begin{equation}
\begin{aligned}
&\sum_{\{i>j,|\mathscr{j}\rangle_i=0,|\mathscr{j}\rangle_j=1\}}(-1)^{\zeta(i,j)}V_2(i,j)
\\&+\sum_{\{i>j,|\mathscr{j}\rangle_i=0,|\mathscr{j}\rangle_j=|\mathscr{j}\rangle_k=1\}}(-1)^{\zeta(i,j)}V_3(i,j,k)
\\&+\sum_{\{i>j>k>l,|\mathscr{j}\rangle_i=|\mathscr{j}\rangle_l=0,|\mathscr{j}\rangle_j=|\mathscr{j}\rangle_k=1\}}(-1)^{\zeta(i,j)+\zeta(k,l)}V_4(i,j,k,l)
\\&+\sum_{\{i>j>k>l,|\mathscr{j}\rangle_i=|\mathscr{j}\rangle_k=0,|\mathscr{j}\rangle_j=|\mathscr{j}\rangle_l=1\}}(-1)^{\zeta(i,j)+\zeta(k,l)}V_5(i,j,k,l)\\&
+\sum_{\{i>j>k>l,|\mathscr{j}\rangle_i=|\mathscr{j}\rangle_j=0,|\mathscr{j}\rangle_k=|\mathscr{j}\rangle_l=1\}}(-1)^{\zeta(i,j)+\zeta(k,l)}V_6(i,j,k,l),
\end{aligned}
\end{equation}
where $\zeta(i,j)$ is the number of `1's that occur in the range $[i,j)$. The upper-triangular elements of $H$ can be obtained by Hermiticity.
$H$ conserves total moir\'e momentum $\bk$ and valley charge, so that we can diagonalize these smaller symmetry sub-blocks to obtain eigenvalues and eigenvectors.

\section{Self-consistent Hartree Fock mean field}
\label{sec:hf}
For the self-consistent HF calculations of $H$, we follow the standard numerical procedure described in e.g.~Refs.~\cite{kwan2021kekule,Kwan2025meanfieldreview}. The HF framework involves searching for the lowest-energy state in the manifold of single Slater determinants. As we are interested in the IKS, we constrain the Slater determinants to preserve a modified translation symmetry with intervalley spiral moir\'e wavevector $\bq$. At the level of the one-body density matrix, this corresponds to the parameterization
\begin{gather}
    \langle c^\dagger_{s,\eta,\bk,m}c_{s',\eta,\bk',n}\rangle=\delta_{s,s'}\delta_{\bk,\bk'}P_{\eta,m;\eta,n}(\bk,s),\\
    \langle c^\dagger_{s,-,\bk,m}c_{s',+,\bk',n}\rangle=\delta_{s,s'}\delta_{\bk+\bq,\bk'}P_{-,m;+,n}(\bk,s).
\end{gather}
Note that intervalley coherence in the HF state is only permitted between $\bk$ in valley $\eta=-$ and $\bk+\bq$ in valley $\eta=+$. If the HF solution corresponds to an IKS, we further enforce it to satisfy $SU(2)_\mathrm{s}$ spin-rotation invariance, as this has been shown to capture the lowest energy state in our present modelling of the TBG Hamiltonian~\cite{wang2025putting}. For an IKS state, we also refer to the intervalley wavevector $\bq$ as the IKS wavevector $\bq_{\text{IKS}}$. The HF ground state is obtained by performing HF calculations with multiple initial seeds for all $\bq$ in the mBZ and minimizing over the total energies. We note that the ordering wavevector is very soft, in the sense that IKS states with considerably different values of $\bq_\text{IKS}$ can have closely competing energy differences of $\lesssim 1\,\text{meV}$ per unit cell~\cite{kwan2021kekule} (see also Fig.~\ref{fig:allq_energy}).

\section{Continuous field quantum Monte Carlo algorithm} 
\label{sec:qmc}

We apply Trotter decomposition to evenly slice the partition function
\begin{equation}
\begin{aligned}
Z=&\Tr\left(\e^{-\beta H}\right)\\
=&\Tr\left(\prod_\tau\e^{-\triangle\tau H}\right)\\
=&\Tr\left(\prod_\tau\left(\prod_\bQ\left(\e^{-\frac{1}{2}\alpha(\bQ)A^2_\bQ}\e^{\frac{1}{2}\alpha(\bQ)B^2_\bQ}\right)\e^{-\triangle\tau H_0}\right)\right)\\&+\mathcal{O}\left(\alpha(\bQ\rightarrow 0)\right),
\end{aligned}
\end{equation}
where $\beta=1/(k_\mathrm{B}T)$ with $k_\mathrm{B}T=0.167$ meV for $L_1\times L_2=18\times18$, while the imaginary time is $\tau\in[0,\beta]$ taken in steps of $\triangle\tau=\beta/600$. We have defined $\alpha(\bQ)=\triangle\tau V(\bQ)/2\Omega$, where $\alpha(\bQ\rightarrow 0)=8\times10^{-3}$ for the simulation parameters used here.
Then, we decouple the interaction terms using the Gaussian integrals
\begin{gather}
\e^{\frac{1}{2}\alpha(\bQ)(iA_\bQ)^2}=\frac{1}{\sqrt{2\pi}}\int^\infty_{-\infty}dx\e^{-\frac{1}{2}x^2}\e^{-ix\sqrt{\alpha(\bQ)}A_\bQ}\\
\e^{\frac{1}{2}\alpha(\bQ)B_\bQ^2}=\frac{1}{\sqrt{2\pi}}\int^\infty_{-\infty}dx\e^{-\frac{1}{2}x^2}\e^{-x\sqrt{\alpha(\bQ)}B_\bQ},
\end{gather}
with the auxiliary field $x$. Therefore, the partition function can be recast as
\begin{equation}
\begin{aligned}
Z=&\int\left(\prod_{\tau,\bQ}dx_{\tau,\bQ,1}dx_{\tau,\bQ,2}\right)\e^{-\frac{1}{2}\sum_{\tau,Q}\left(x^2_{\tau,\bQ,1}+x^2_{\tau,\bQ,2}\right)}\\&\times\Tr\left(\prod_\tau\e^{i\sum_\bQ\left(-x_{\tau,\bQ,1}\sqrt{\alpha(\bQ)}A_\bQ+i x_{\tau,\bQ,2}\sqrt{\alpha(\bQ)}B_\bQ\right)}\e^{-\triangle\tau H_0}\right)\\
&+\mathcal{O}(8\times10^{-3}).
\end{aligned}
\label{eq:zcfmc}
\end{equation}
Note that the decoupling is exact, so that the error in $Z$ above comes from the Trotter decomposition.

We apply Hamiltonian dynamics to update $x$. Rewrite the partition function as 
\begin{equation}
\begin{aligned}
Z=&\int\left(\prod_{i}dx_{i}d p_{i}\right)\e^{-\frac{1}{2}\sum_{i}\left(x^2_{i}+p^2_{i}\right)}\det\left(\I+B_C(\beta,0)\right)\\
&+\mathcal{O}(8\times10^{-3}),
\end{aligned}
\end{equation}
where the composite index $i\equiv({\tau,\bQ,n})$ with $n=1,2$ is employed for compactness and an artificial momentum $p_{i}$ is introduced for each $x_{i}$. $B_C(\beta,0)$ is the matrix form of $\prod_\tau\e^{i\sum_\bQ\left(-x_{\tau,\bQ,1}\sqrt{\alpha(\bQ)}A_\bQ+i x_{\tau,\bQ,2}\sqrt{\alpha(\bQ)}B_\bQ\right)}\e^{-\triangle\tau H_0}$. $C$ stands for a configuration of all the auxiliary fields $\{x_{i}, p_{i}\}$. The partition function can be further expressed as
\begin{equation}
\begin{aligned}
Z=&\int\left(\prod_{i}dx_{i}d p_{i}\right)\e^{-\mathcal{ H}}+\mathcal{O}(8\times10^{-3}),
\end{aligned}
\end{equation}
with the effective Hamiltonian 
\begin{equation}
\mathcal{H}\equiv\frac{1}{2}\sum_{i}\left(x^2_{i}+p^2_{i}\right)-\ln \left(\det\left(\I+B_C(\beta,0)\right)\right).
\end{equation}

The Hamiltonian dynamics is applied as  
\begin{equation}
\begin{aligned}
\frac{dp_{i}}{dt}=&-\frac{\partial\mathcal{H}}{\partial x_{i}}\\
= &-x_{i}+2\sum_\eta\Tr\left(J^\eta_\bQ(\I-G^\eta(\tau))\right) +\mathcal{O}(10^{-2}) \\
\frac{dx_{i}}{dt}=&\frac{\partial\mathcal{H}}{\partial p_{i}}\\
=&p_{i},
\label{eq:ham_dyn}
\end{aligned}
\end{equation}
where $G^\eta(\tau)$ is the equal-time Green's function for $s=\uparrow$, and $J^\eta_\bQ$ is the matrix form of $A^\eta_\bQ$ or $B^\eta_\bQ$  for $n = $ 1 or 2 respectively and is also for $s=\uparrow$. The Hamiltonian dynamics facilitates increased acceptance rates through the conservation of the Hamiltonian along equal-energy trajectories. The Metropolis-Hastings scheme is used to determine acceptance or rejection of the update of $x_i$.

\section{Approximated quantum Monte Carlo}
\label{sec:AQMC}
The partition function can be further written as
\begin{equation}
\begin{aligned}
Z=&\int dCW_C,
\end{aligned}
\end{equation}
where $W_C$ is generally a complex number in a configuration $C$. An observable $\hat{O}$ can be measured by
\begin{equation}\label{eq:QMC_sampling}
\begin{aligned}
\langle \hat{O}\rangle&=\frac{\int dCO_CW_C}{\int dCW_C}
=\frac{\int dC O_C \frac{W_C}{\mathrm{Re}(W_C)} \,\mathrm{Re}(W_C) } {\int dC \mathrm{Re}(W_C)}
\end{aligned}
\end{equation}
where $O_C$ is the value of $\hat{O}$ in $C$. The reality of the partition function leads to $\int dC W_C=\int dC \text{Re}(W_C)$. 
Therefore, performing importance sampling over the auxiliary field with the sampling weight $\mathrm{Re}(W_C)$ yields  the  expectation value $\langle \hat{O}\rangle$. 

Away from charge neutrality ($\nu=0$), $\mathrm{Re}(W_C)$  can be negative~\cite{zhangMomentum2021,zhangPolynomial2023,huangEvolution2024,huangAngle2025}, which introduces significant difficulty in obtaining reliable statistics. To circumvent this problem, we make an approximation by replacing the true sampling weight in Eq.~\eqref{eq:QMC_sampling} by its absolute magnitude $|\text{Re}(W_C)|$ which corresponds to a modified statistical ensemble. In other words, we evaluate $\langle \hat{O}\rangle$ according to
\begin{equation}
\begin{aligned}\label{eq:QMC_sampling_approximate}
\langle \hat{O}\rangle&\leftarrow \frac{\int dC O_C \frac{W_C}{\mathrm{Re}(W_C)} \,  |\mathrm{Re}(W_C)|}{\int dC |\mathrm{Re}(W_C)|}.
\end{aligned}
\end{equation}
In Appendix~\ref{sec:property_aqmc}, we prove that the approximated treatment shares the same symmetries as the original Hamiltonian, and therefore does not introduce explicit breakings in the symmetries investigated. 

\section{Properties of Approximated QMC}
\label{sec:property_aqmc}
\label{sec:pAQMC}
We consider the Hamiltonian 
\begin{equation}
\begin{aligned}
    \hat{H}=&\sum_{s,\eta,\bk,m}\left(\epsilon^{s,\eta}_{\bk,m}-\mu\right)c^\dagger_{s,\eta,\bk,m} c_{s,\eta,\bk,m}\\
    &+\sum_{\bQ>0}\frac{1}{4\Omega}V(\bQ)\left(A^2_\bQ-B^2_\bQ\right),
    \label{eq:H}
\end{aligned}
\end{equation}
which generally possesses a sign problem in the auxiliary field QMC method.
The  question we address here is whether there is a proximate Hamiltonian  that shares the same  symmetries of  those of the original Hamiltonian,  but is free of the negative sign problem by construction.   As discussed above, by using a Gaussian transformation to  decouple the  interaction term, we show that the  partition function takes the form
\begin{equation}
\begin{aligned}
Z=&\int dC\,\text{Re}(W_C),
\end{aligned}
\end{equation}
with
\begin{gather}
W_C=\e^{-S_B(C)}\Tr\left(\prod_\tau\e^{\hat{h}(C_\tau)}\right)\\
S_{B} (C)   = \frac{1}{2}\sum_{i}\left(x^2_{i}+p^2_{i}\right)\\
\hat{h}(C_\tau)=i\sum_\bQ\left(-x_{\tau,\bQ,1}\sqrt{\alpha(\bQ)}A_\bQ+i x_{\tau,\bQ,2}\sqrt{\alpha(\bQ)}B_\bQ\right)-\triangle\tau H_0.
\end{gather}
$C_\tau$ denotes the auxiliary field configuration at a given Euclidean time $\tau$. We note that observables  are  given by
\begin{equation}\label{eq:O_EV}
\langle \hat{{O}} \rangle = \frac{1}{Z} \int dC \,\text{Re} (W_C) \left[\frac{W_C}{\text{Re} W_C} O_C\right]
\end{equation}
with
\begin{equation}
O_C =  
\frac{\Tr\left(\prod_\tau\e^{\hat{h}(C_\tau)}\hat{O}\right)}{\Tr\left(\prod_\tau\e^{\hat{h}(C_\tau)}\right)}.
\end{equation}
$\hat{h}(C_\tau)$  does not  necessarily have the  symmetries of the original Hamiltonian,    but after integration over  field configurations
they must be restored since the Gaussian decoupling is exact.  For example, in our specific case, $\hat{h}(C_\tau)$   enjoys  the same  spin-$SU(2)$ and valley-$U(1)$ symmetries as the original  Hamiltonian, but the
moir\'e translation symmetry  $\hat{T}_{\mathbf{R}}$ is broken.  In particular, the operators in $\hat{h}(C_\tau)$ transform as   
\begin{gather}\label{eq:AQ_BQ_transform}
\hat{T}^{-1}_{\mathbf{R}}A_\bQ\hat{T}_{\mathbf{R}} 
= \cos(\bQ\cdot\bR)A_\bQ+i\sin(\bQ\cdot\bR)B_\bQ\\
\hat{T}^{-1}_{\mathbf{R}}B_\bQ\hat{T}_{\mathbf{R}} 
= \cos(\bQ\cdot\bR)B_\bQ+i\sin(\bQ\cdot\bR)A_\bQ.
\end{gather}     
However, translation symmetry will be  restored after summation over the field. To see this explicitly from Eq.~\ref{eq:O_EV}, recall that the presence of a symmetry $\hat{U}$ (i.e.~satisfying $[\hat{U},\hat{H}]=0$) implies that expectation values satisfy $\langle \hat{U}\hat{O}\hat{U}^\dagger\rangle=\langle\hat{O}\rangle$. Taking $\hat{O}\rightarrow \hat{U}\hat{O}\hat{U}^\dagger$ in $O_C$ in Eq.~\ref{eq:O_EV} induces an effective orthogonal transformation in the auxiliary fields in $\hat{h}(C_\tau)$ of the numerator according to
\begin{gather}
    x_{\tau,Q,1}\rightarrow\cos(\bQ\cdot\bR)x_{\tau,Q,1} + \sin(\bQ\cdot\bR)x_{\tau,Q,2}\\
        x_{\tau,Q,2}\rightarrow\cos(\bQ\cdot\bR)x_{\tau,Q,2}-\sin(\bQ\cdot\bR)x_{\tau,Q,1}.
\end{gather}
Denote this transformation as $\underline{T}$, i.e. $C\rightarrow \underline{T}C$. Owing to the cyclicity of the trace, we have that the denominator of $O_C$ is invariant under $C\rightarrow \underline{T}C$, from which we conclude that $O_C\rightarrow O_{\underline{T}C}$ when considering the expectation value of $ \hat{U}\hat{O}\hat{U}^\dagger$. Since $S_B(C)$ is invariant under orthogonal transformations, we also find that $W_C=W_{\underline{T}C}$ by cyclicity of the trace. We can therefore shift integration variables $C\rightarrow \underline{T}^{-1}C$ (with unit Jacobian) in the expression for $\langle \hat{U}\hat{O}\hat{U}^\dagger\rangle$, demonstrating that $\langle \hat{U}\hat{O}\hat{U}^\dagger\rangle=\langle\hat{O}\rangle$. We can perform a similar analysis for $C_2$ which transforms $A_\bQ\rightarrow A_\bQ$ and $B_\bQ\rightarrow -B_\bQ$, and time reversal $\mathcal{T}$, to show that they are symmetries captured by the measurement of $\langle\hat{O}\rangle$ (despite not being symmetries of $\hat{h}(C_\tau)$).

The  sign  problem arises from the fact that 
$\text{Re} (W_C)$ can be negative for some  fields.   While it is possible to cope  with the sign problem  if it is not too severe,  one  can 
ask the question whether  sampling the partition function 
\begin{equation}
      \overline{Z} = \int dC \left| \text{Re}(W_C)\right|,
\end{equation}
which by construction does not have a sign problem, corresponds to a  Hamiltonian  $\overline{\hat{H}}$ with the same symmetries as the original one.   
The  symmetry properties of 
 $\overline{\hat{H}}$ depend strongly upon  the choice of  the Hubbard-Stratonovich (HS) transformation.   For  our example, $\overline{\hat{H}}$   continues to obey all the symmetries because taking $\text{Re}(W_C)\rightarrow |\text{Re}(W_C)|$ does not interfere with the property $W_C=W_{\underline{T}C}$ used above.

We  now examine whether $\overline{\hat{H}}$  is a  Hermitian operator, which would imply that the density matrix $\bar{\hat{\rho}} = e^{-\beta \overline{\hat{H}}}$  is  a  positive semi-definite Hermitian operator.   
In the  \textit{high} temperature limit  where the sign problem is not severe,  we  expect $\overline{\hat{\rho}}$  to be  proximate  to $\hat{\rho}$.   At low temperatures,  the  notion of proximate  can be tested explicitly on case-by-case basis by e.g.~comparison to  exact diagonalization.

 The  auxiliary field  QMC  method  can  be viewed as a  sampling of the density matrix $\hat{\rho}$   in the space of Gaussian operators.  In particular,  we can follow Ref.~\cite{GroverEntanglement2013} to  write 
\begin{equation}
    \hat{\rho} = \frac{\int dC  W(C)   \hat{\rho}(C)}{\int dC  W(C) }
\end{equation}
with
\begin{equation}
\hat{\rho}(C) =  \det{G(C)} e^{-\hat{c}^{\dagger} \ln{(G^{-1}(C)-1)^{-1}} \hat{c}} 
\end{equation}
and
\begin{equation}
G(C)_{x,y} =\frac{\Tr\left(\prod_\tau\e^{\hat{h}(C_\tau)}\hat{c}^{\phantom\dagger}_{x} \hat{c}^{\dagger}_{y}\right)}{\Tr\left(\prod_\tau\e^{\hat{h}(C_\tau)}\right)}. 
\end{equation}
Since the  density matrix is Hermitian,  we can again sample: 
\begin{equation}
        \hat{\rho} = \frac{\int dC  \text{Re} W(C)  \frac{1}{2 \text{Re} W(C)}\left(  W(C) \hat{\rho}(C) + W^*(C) \hat{\rho}^\dagger(C) \right) }{\int dC  \text{Re}\left(W(C)\right) }.
\end{equation}
The  density matrix  associated to  $\overline{\hat{H}}$  will then be given by: 
\begin{equation}
\overline{\hat{\rho}} = \frac{\int dC  \left| \text{Re}\left(W(C)\right) \right| \frac{1}{2 \text{Re} W(C) }\left(  W(C) \hat{\rho}(C) + W^*(C) \hat{\rho}^\dagger(C) \right) }{\int dC  \left| \text{Re}\left(W(C)\right) \right| }. 
\end{equation}
$\overline{\hat{\rho}}$  is  thus Hermitian by construction but  not   necessarily   positive  such  that  $\overline{\hat{H}}$   could be non-Hermitian   with 
complex eigenvalues. The imaginary part of the  eigenvalue is however quantized  to $m\pi/\beta$ due to  the  Hermiticity of $\overline{\hat{\rho}}$. The particular values of $m$ depend on the model and the applied HS decomposition. We defer a more detailed study of these matters in an upcoming paper.  For now, we comment that test calculations on small clusters suggest that the weights of negative eigenvalues of the density matrix tend to be stable in the low-$T$ limit, and disappear in the high-$T$ limit, where the sign problem gradually goes away. 

\begin{figure}[!ht]
\centering
\includegraphics[width=\linewidth]{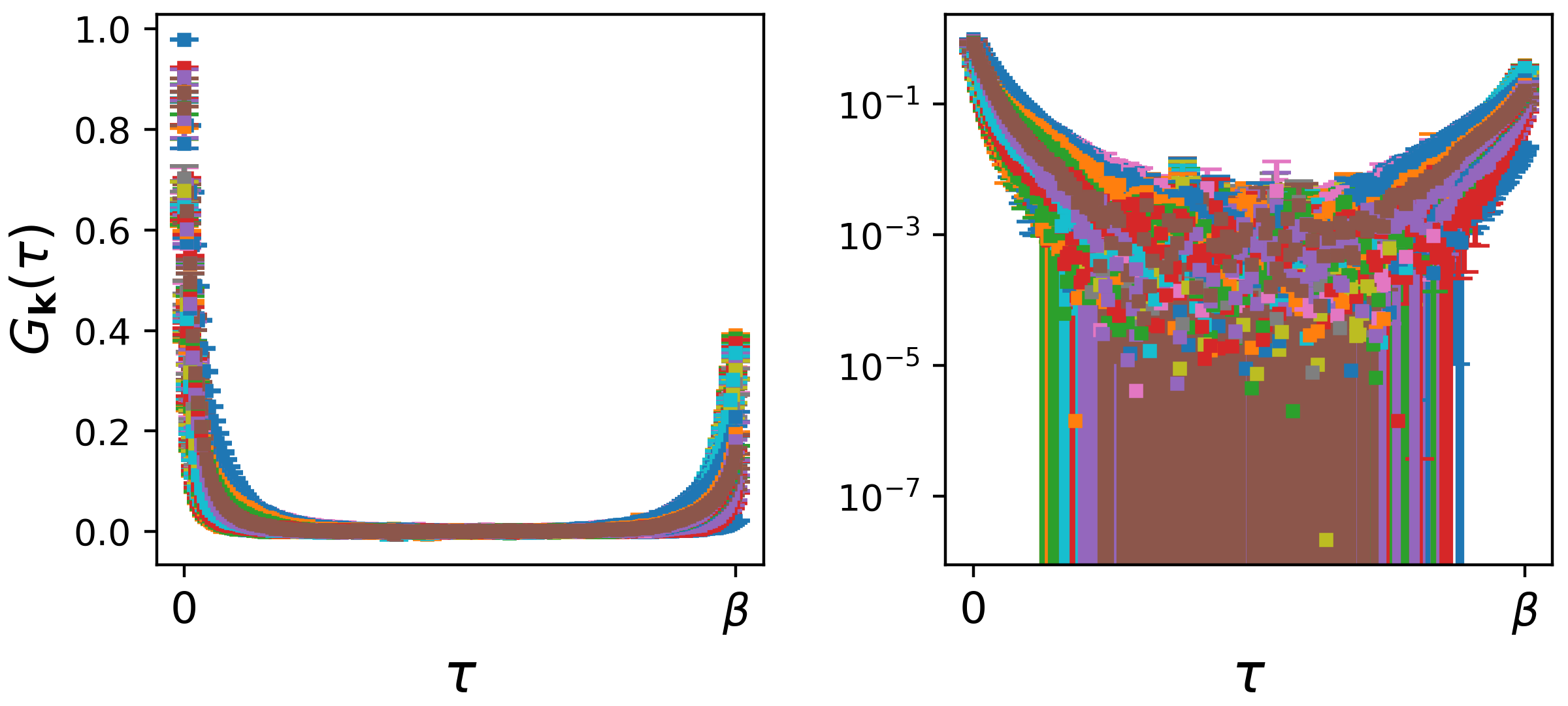}
\caption{Green's functions at the strains of $\epsilon_\s=$ 0.6\% for $L_1\times L_2=6\times6$ within AQMC. In left panel y-axis is in linear scale while the right panel is in log scale. Different colors refer to different $\bk$ points. All the Green functions are gapped. Model parameters are $\theta=1.15^\circ,\phi_\s=130^\circ, u_0=80\,$meV and $\epsilon_\r=10$.}
\label{fig:Gt0_L6_strn6}
\end{figure}

\begin{figure}[!ht]
\centering
\includegraphics[width=\linewidth]{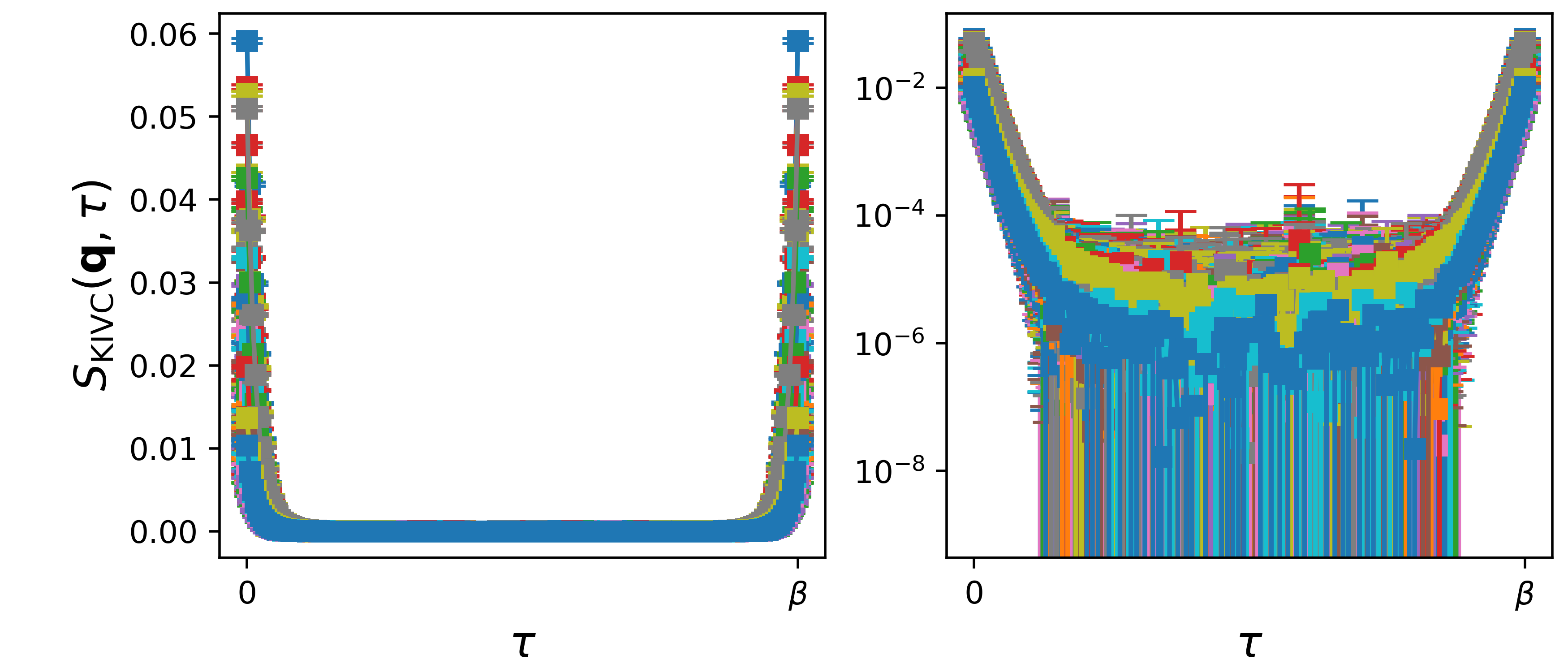}
\caption{KIVC structure factor at $\epsilon_\s=$ 0.6\% for $L_1\times L_2=6\times6$ within AQMC. In left panel y-axis is in linear scale while the right panel is in log scale. Different colors refer to different $\bq$ points. Model parameters are $\theta=1.15^\circ,\phi_\s=130^\circ, u_0=80\,$meV and $\epsilon_\r=10$.}
\label{fig:Sqt_L6_strn6_KIVC}
\end{figure}

\begin{figure}[!ht]
\centering
\includegraphics[width=\linewidth]{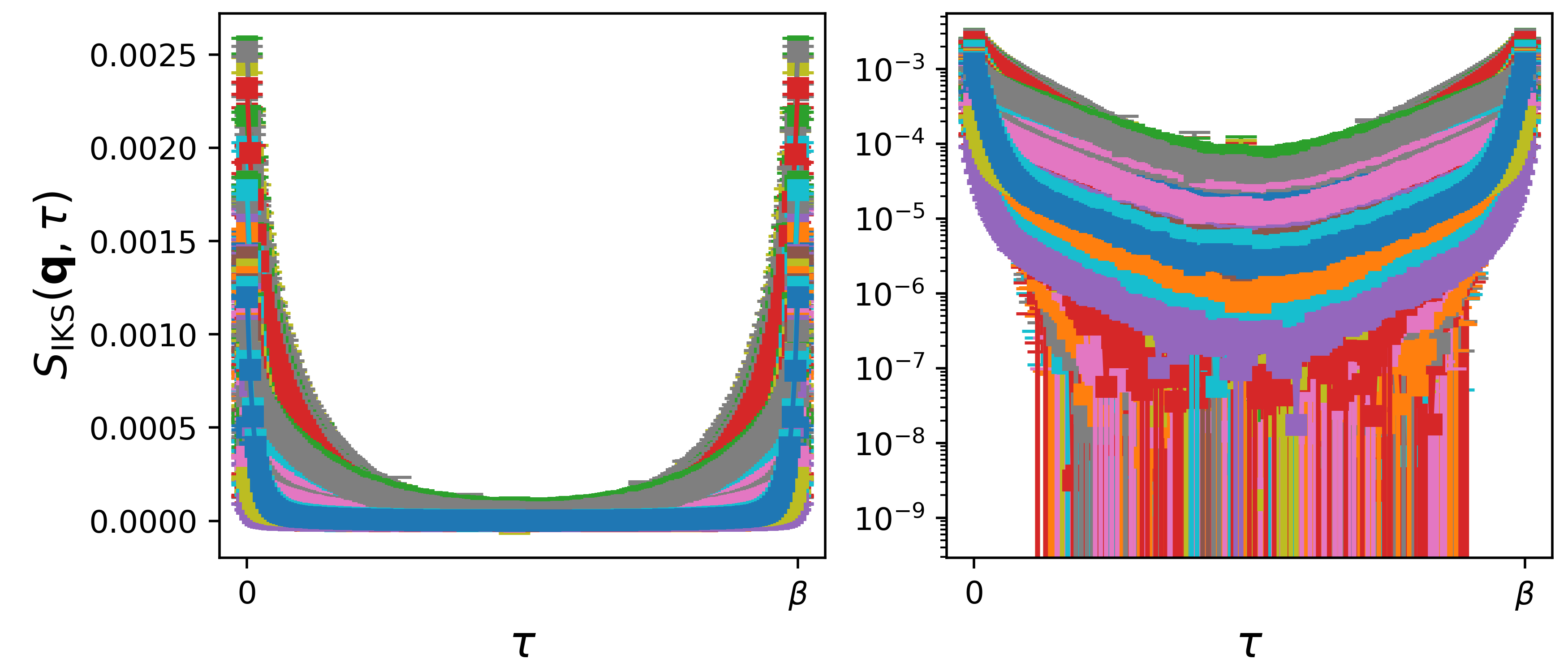}
\caption{IKS structure factor at $\epsilon_\s=$ 0.6\% for $L_1\times L_2=6\times6$ within AQMC. In left panel y-axis is in linear scale while the right panel is in log scale. Different colors refer to different $\bq$ points. Model parameters are $\theta=1.15^\circ,\phi_\s=130^\circ, u_0=80\,$meV and $\epsilon_\r=10$.}
\label{fig:Sqt_L6_strn6_IKS}
\end{figure}

Another way to check the presence of non-Hermiticity of the effective Hamiltonian is to look for oscillatory behavior of the Euclidean time-displaced correlators. If a complex eigenvalue $\lambda_\mathcal{C}$ arises, then Euclidean time displaced  correlation functions will have a  damped oscillatory behavior with  period  $2 \pi/ \mathrm{Im} (\lambda_\mathcal{C}) $. 
We have attempted to detect such oscillatory behavior in the single-particle Green's function, the KIVC structure factor and the IKS structure factor, as shown in Fig.~\ref{fig:Gt0_L6_strn6}, Fig.~\ref{fig:Sqt_L6_strn6_KIVC} and Fig.~\ref{fig:Sqt_L6_strn6_IKS}. However, since $\mathrm{Im} (\lambda_\mathcal{C}) =  m\pi/\beta$ the period is $2 \beta/m$, and would be hard to  detect for  small values of $m$. There is also a possibility of our probes being orthogonal to the negative eigenvector of the effective density matrix $\overline{\hat{\rho}}$. Hence we can only conclude that non-Hermiticity of the effective Hamiltonian   $\overline{\hat{H}}$  does not obviously manifest itself in single-particle or KIVC (IKS) structure factors.

Despite the rather different mathematical formulation, we propose that our approximated QMC shares similarities in physical meaning  to the dominant thimble approximation in Ref.~\cite{ulybyshevBeyond2024}. In the latter approximation, we neglect the fluctuations of the sign of $W_C$, because the Monte Carlo field configurations are on a thimble (which is by definition a manifold with fixed sign factor), and we choose the thimble with the maximal weight in the full sum over thimbles forming a partition function. Physically, it means that the saddle point for this thimble captures the correct mean-field solution, and the surrounding thimble gives the fluctuations around it, which can be described neglecting the fluctuations of the sign due to aforementioned mathematical definition of a thimble. Thus, in the approximated QMC, if we are sufficiently close to the mean field solution, similar to the dominant thimble approximation, the fluctuation of the sign might play a lesser role in the description of the fluctuations around this mean field Slater determinant state.

\section{More benckmarks among QMC, AQMC and ED}
\label{sec:Benchmark_ED_AQMC}
\begin{figure}[!ht]
\centering
\includegraphics[width=\linewidth]{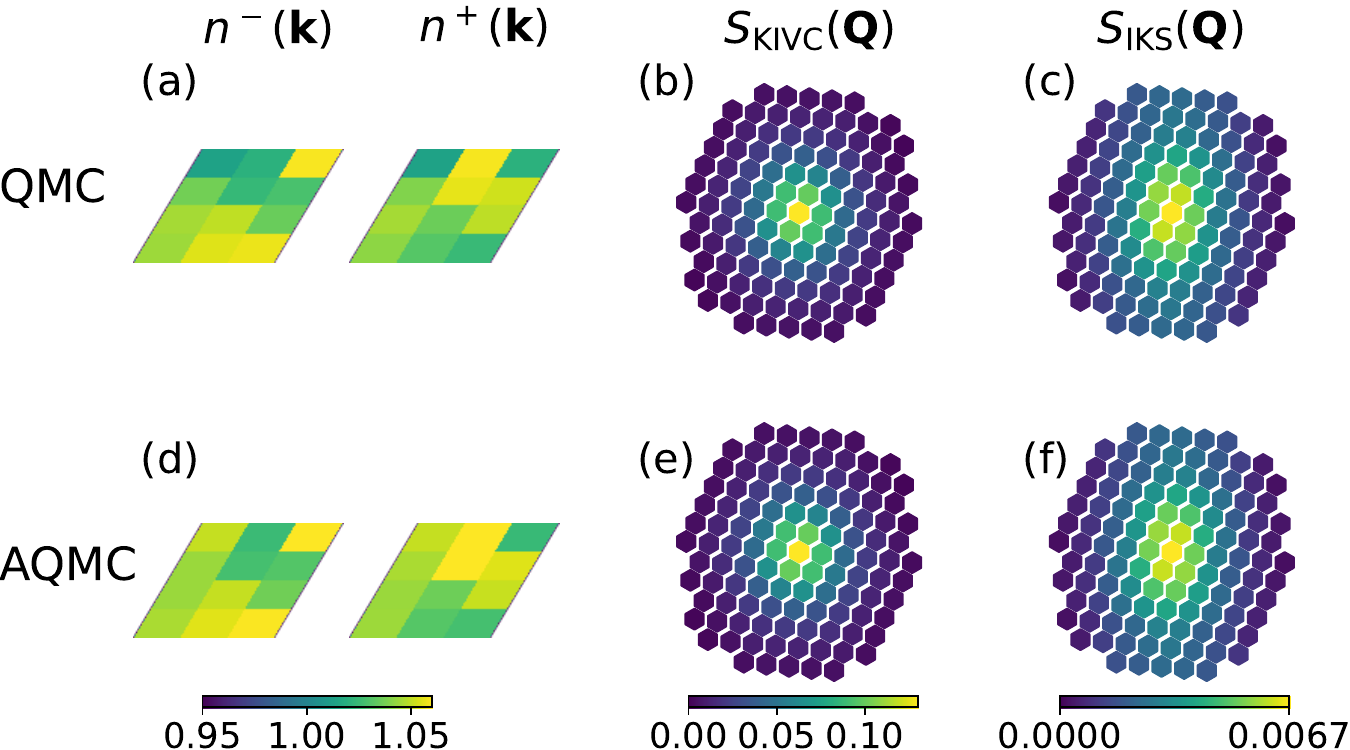}
\caption{Comparison of occupation number $n^\eta(\bk)$ ((a) and (d)) and structure factors ($S_\mathrm{KIVC} (\bQ)$ ((b) and (e)) for KIVC  and $S_\mathrm{IKS}(\bQ)$ ((c) and (f)) for IKS) from stringent quantum Monte Carlo (QMC) and approximated quantum Monte Carlo (AQMC) with $\epsilon_\s=$ 0.3\% and $L_1\times L_2=4\times3$. We consider spinful $\nu=-2$. Note that the parameters are $\theta=1.15^\circ,\phi_\s=130^\circ, u_0=80\,$meV and $\epsilon_\r=10$. The average sign for stringent QMC is $\langle\mathrm{sgn}\rangle=0.649$.}
\label{fig:QMC-AQMC}
\end{figure}

For a high temperature of $k_\mathrm{B}T=150$ meV, we also provide the benchmarks between QMC and AQMC for the spinful system of $L_1\times L_2=4\times3$ in Fig.~\ref{fig:QMC-AQMC}, where the observables match well.

\begin{figure}[!ht]
\centering
\includegraphics[width=\linewidth]{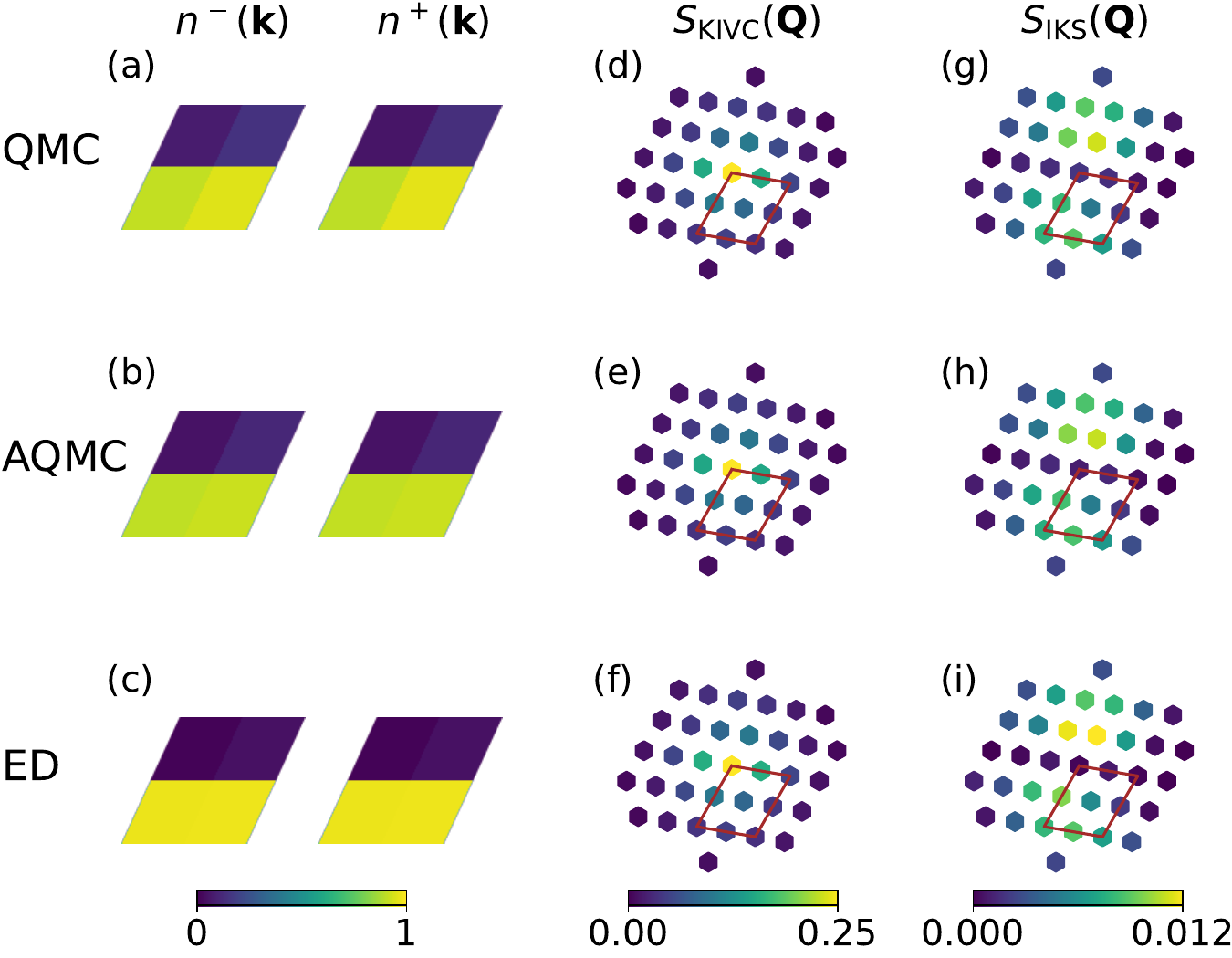}
\caption{Comparison of occupation number $n^\eta(\bk)$ ((a)-(c)) and structure factors ($S_\mathrm{KIVC}(\bQ)$ ((d)-(f)) for KIVC  and $S_\mathrm{IKS}(\bQ)$ ((g)-(i)) for IKS) from stringent quantum Monte Carlo (QMC), approximated quantum Monte Carlo (AQMC), and exact diagonalization (ED) with $\epsilon_\s=$ 0.6\% and $L_1\times L_2=2\times2$ . We consider spinless calculations for $\nu=-1$, which is related to spinful $\nu=-2$. Note that the parameters are $\theta=1.15^\circ,\phi_\s=130^\circ, u_0=80\,$meV and $\epsilon_\r=10$. The average sign for stringent QMC is $\langle\mathrm{sgn}\rangle=0.043$. The mBZ in (d)-(i) is denoted by the brown rhombus.}
\label{fig:QMC-AQMC-ED-L2}
\end{figure}

For the ground state, the benchmarks of the occupation distribution $n^\eta(\bk)$ and IVC structure factors $S_\mathrm{KIVC}$ and $S_\mathrm{IKS}$ among QMC, AQMC and ED for $L_1\times L_2=2\times2$ with the strain strength of $\epsilon_\s=0.6$\% are provided in Fig.~\ref{fig:QMC-AQMC-ED-L2}. As expected, the stringent QMC (with the exponential sign problem) and ED calculations agree with each other. We also find qualitative agreement for $n^\eta(\bk)$, $S_\text{KIVC}(\bq)$ and $S_\text{IKS}(\bq)$ between AQMC and the other exact methods.

\section{Intervalley structure factors}
\label{sec:SI_IVC}

\begin{figure*}[!ht]
\centering
\includegraphics[width=\linewidth]{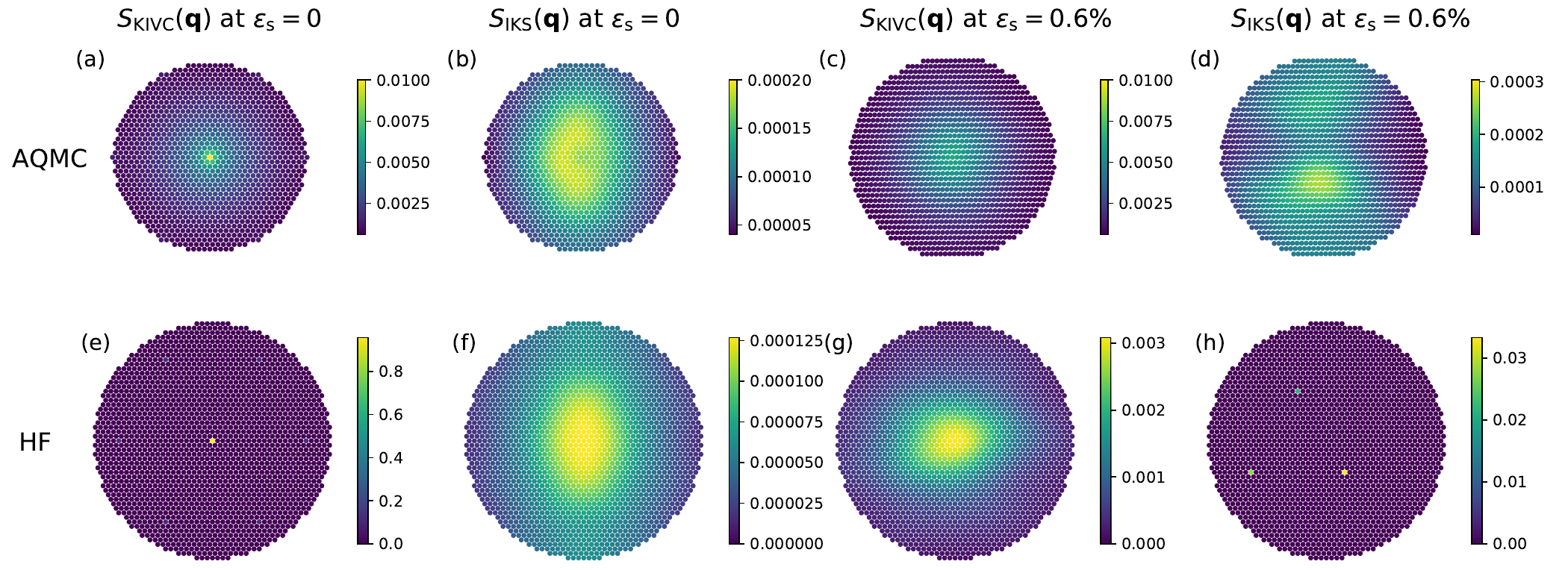}
\caption{Comparisons of KIVC and IKS structure factors ($S_\mathrm{KIVC}(\bq)$ and $S_\mathrm{IKS}(\bq)$) from approximated quantum Monte Carlo (AQMC) and Hartree Fock (HF) at $\nu=-2$ with $N_{\bk}=18\times18$. Upper panels are from AQMC while the lower ones are from HF. $S_\mathrm{KIVC}(\bq)$ with zero strain is shown in (a) and (e), and with $0.6$\% is shown in (c) and (g). $S_\mathrm{IKS}(\bq)$ with zero strain is shown in (b) and (f), and with $0.6$\% is shown in (d) and (h). $\bq=0$ is located at the center of each plot. 
Model parameters are $\theta=1.15^\circ, \phi_\s=130^\circ, u_0=80\,$meV and $\epsilon_\r=10$.}
\label{fig:Sq-QMC-HF-strn0-strn6}
\end{figure*}

In this section we show the comparisons of the KIVC structure factor $S_\mathrm{KIVC}$ and IKS structure factor $S_\mathrm{IKS}$ from both AQMC and HF at $\epsilon_\s=$ 0 and $\epsilon_\s=$ 0.6\% in Fig.~\ref{fig:Sq-QMC-HF-strn0-strn6}. The definitions of these structure factors are given in the main text.

In QMC, a KIVC peak at $\bq=0$ is detected in Fig.~\ref{fig:Sq-QMC-HF-strn0-strn6} (a) at zero strain, while the IKS structure factor does not exhibit a peak as shown in Fig.~\ref{fig:Sq-QMC-HF-strn0-strn6} (b). When a strain of $\epsilon_\s=$ 0.6\% is applied, the KIVC order vanishes as no peak appears in (c), while $S_{\text{IKS}}(\bq)$ does not exhibit a peak in (d) that survives for increasing $L$.

In HF, a strong KIVC peak is seen at $\bq=0$ without strain in Fig.~\ref{fig:Sq-QMC-HF-strn0-strn6} (e), while strong IKS peaks at $\bq=\bq_\text{IKS}+\bG$ are resolved at $\epsilon_\s=$ 0.6\% in Fig.~\ref{fig:Sq-QMC-HF-strn0-strn6} (h). At zero strain (finite strain $\epsilon_\s=$ 0.6\%), no IKS (KIVC) order is observed.

\section{Additional Hartree-Fock Results}
\label{sec:SI_Hartree_Fock_allq}

\begin{figure}[!h]
\centering
\includegraphics[width=0.7\linewidth]{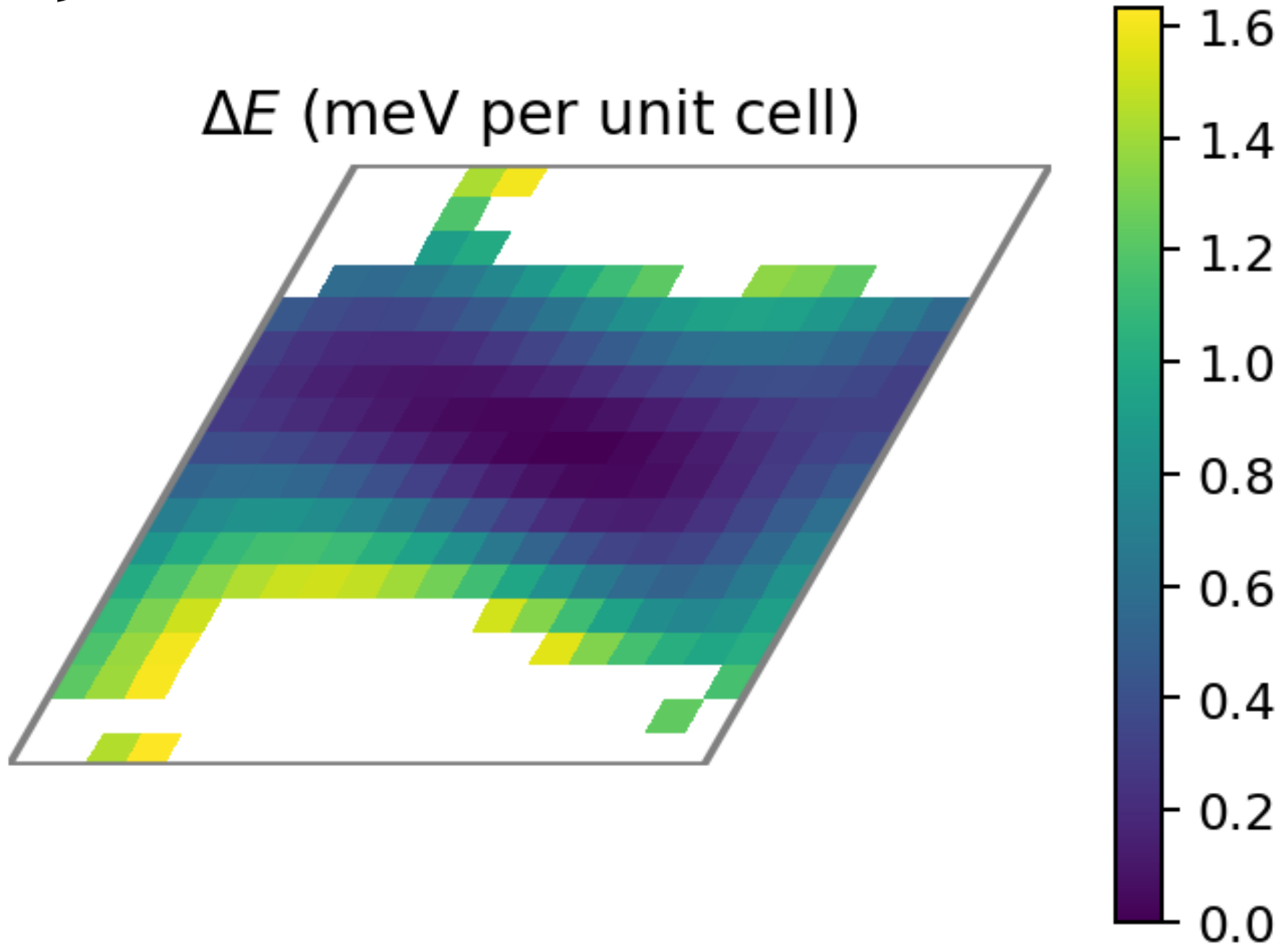}
\caption{HF energy of IKS states as a function of $\bq_\text{IKS}$ at $\nu=-2$ with $N_{\bk}=18\times18$ and $\epsilon_\mathrm{s} = $ 0.3\%. The energy $\Delta E$ is measured relative to the ground state taken across all $\bq_\text{IKS}$. White regions indicate values of $\bq$ where HF did not obtain a time-reversal symmetric IKS. Model parameters are $\theta=1.15^\circ, \phi_\s=130^\circ, u_0=80\,$meV and $\epsilon_\r=10$.}
\label{fig:allq_energy}
\end{figure}

In Fig.~\ref{fig:allq_energy}, we perform HF calculations with the parameters used in Fig.~\ref{fig:nk-Oq-strn0-strn3} for $\epsilon_\s$ = 0.3\%. Recall that in the HF procedure, we perform calculations for all possible values of intervalley spiral moir\'e wavevector $\mathbf{q}$ lying on the momentum mesh. For a large portion of such $\mathbf{q}$, the lowest energy HF state corresponds to an IKS state. In Fig.~\ref{fig:allq_energy}, we show the energy of these IKS states as a function of $\mathbf{q}_\text{IKS}$, measured relative to the global ground state IKS (i.e.~the IKS with the lowest energy across all possible $\mathbf{q}_\text{IKS}$). For the strain of $\epsilon_\s=0.3\%$, we find that the landscape of IKS states is energetically very soft. For example, IKS states with an $\mathbf{q}_\text{IKS}$ that differs by half a RLV compared to the value of $\mathbf{q}_\text{IKS}$ in the global ground state can still be within 0.5\,meV per unit cell. We also find that the region of $\mathbf{q}_{\text{IKS}}$ with low-lying IKS states qualitatively resembles the region where $O''(\mathbf{q})$ is small in the AQMC calculation of Fig.~\ref{fig:nk-Oq-strn0-strn3} (l). The close competition between many different IKS states could be relevant for the inability for AQMC to resolve a long-range-ordered IKS, at least for these parameters, since the softness of the energy for different $\bq_{\mathrm{IKS}}$ points to strong fluctuations around the mean-field ground state.

\section{Single-particle Green's functions}
\label{sec:SIII}
\begin{figure}
\centering
\includegraphics[width=\linewidth]{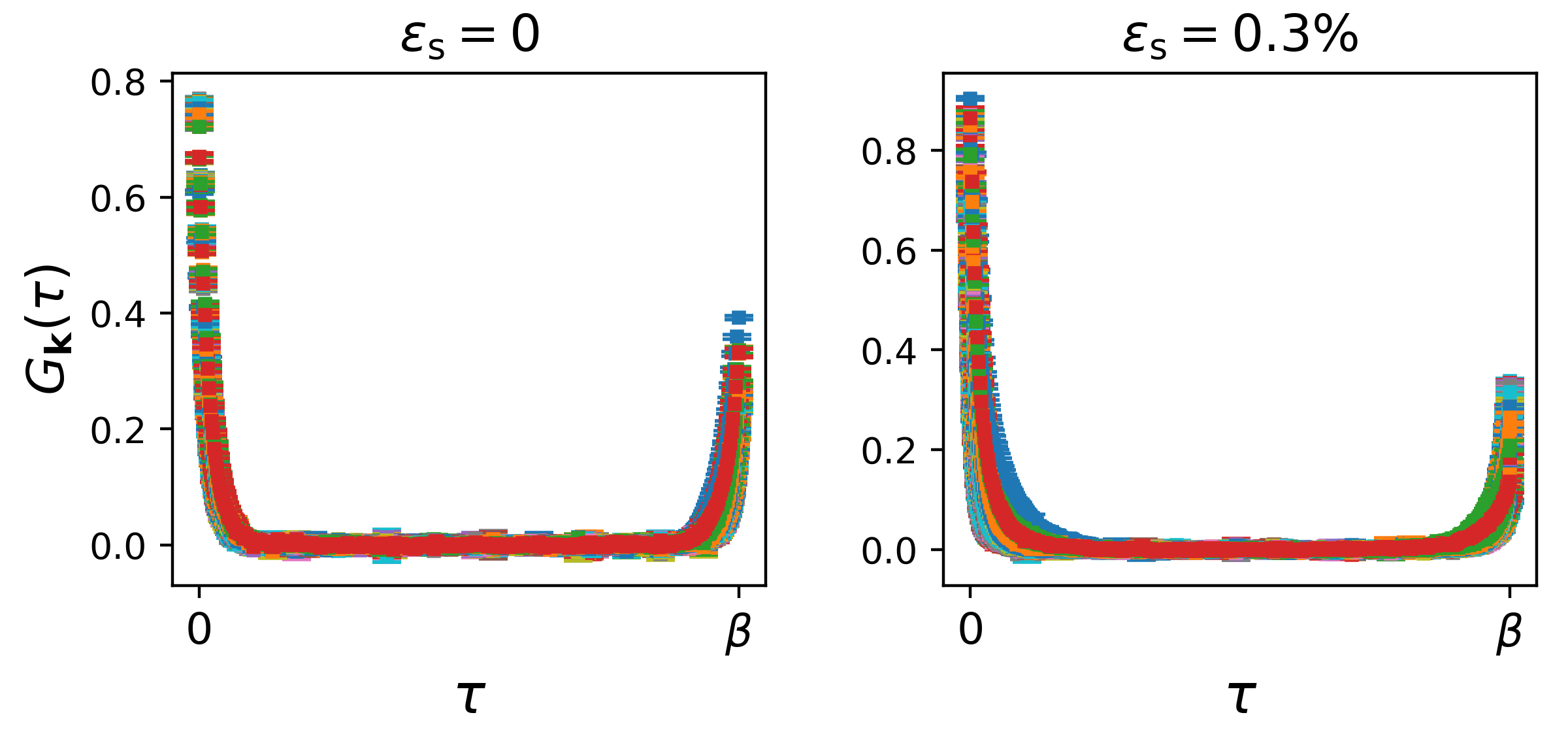}
\caption{Single-particle Green's functions at the strains of $\epsilon_\s=$ 0 and 0.3\% for $L_1\times L_2=12\times12$ within AQMC. Different colors refer to different $\bk$ points. All the Green's functions decay to 0. Model parameters are $\theta=1.15^\circ,\phi_\s=130^\circ, u_0=80\,$meV and $\epsilon_\r=10$.}
\label{fig:Gt0}
\end{figure}

In this section, we show that the interacting system for $\epsilon_s=0.0\%,0.3\%$ in AQMC is gapped by monitoring the imaginary time single-particle Green's function in Fig.~\ref{fig:Gt0}. The Green's functions  were calculated for the system size of $L_1\times L_2=12\times12$  and the temperature of $1/\beta=0.25$ meV. For all the $\bk$ points in the mBZ, the $G_{\mathbf{k}}(\tau)$ all decay quickly to 0 as a function of imaginary time for  both strains. Such decay implies that the single-particle spectrum is gapped at all the momenta considered and hence the system is an insulator from the electronic perspective.

\bibliography{refs}

\end{document}